\documentstyle[12pt,prd,aps,psfig,floats]{revtex}
\tightenlines
\def\lsim{\mathrel{\rlap{\lower4pt\hbox{\hskip1pt$\sim$}}
    \raise1pt\hbox{$<$}}}         
\def\gsim{\mathrel{\rlap{\lower4pt\hbox{\hskip1pt$\sim$}}
    \raise1pt\hbox{$>$}}}         
\def\freg{f_{\mbox{\rm reg}}}    
\begin{document}
\begin{titlepage}

\preprint{
\vbox{
         \hbox{IFIC/00-40}
         \hbox{IASSNS-HEP-00-51}
         \hbox{hep-ph/0007227}
         \hbox{}\hbox{} }}

\title{Phenomenology of Maximal and Near--Maximal Lepton Mixing}

\vskip 1cm

\author{M. C. Gonzalez-Garcia \thanks{concha@flamenco.ific.uv.es}
and Carlos Pe\~na-Garay \thanks{penya@flamenco.ific.uv.es}}
\address{Instituto de F\'{\i}sica Corpuscular \\
Universitat de  Val\`encia -- C.S.I.C\\
Edificio Institutos de Paterna, Apt 22085, 46071 Val\`encia, Spain}
\vskip 1cm
\author{{Yosef Nir}\thanks{ftnir@wicc.weizmann.ac.il}} 
\address{School of Natural Sciences, 
Institute for Advanced Study, Princeton, NJ 08540, USA\\
Department of Particle Physics, Weizmann Institute of Science,
 Rehovot 76100, Israel} 
\vskip 1cm
\author{Alexei Yu. Smirnov\thanks{smirnov@ictp.trieste.it}}
\address{
International Center for Theoretical Physics, 34100 Trieste, Italy\\
Institute for Nuclear Research, Russian Academy of Sciences, 
Moscow 117312, Russia} 
\maketitle

\vskip 0.5cm

\begin{abstract}
The possible existence of maximal or near--maximal lepton
mixing constitutes an intriguing challenge for fundamental theories
of flavour. We study the phenomenological consequences of maximal  
and near--maximal mixing of the electron neutrino with other 
($x$=tau and/or muon) neutrinos. We describe the deviations from 
maximal mixing in terms of a parameter 
$\epsilon\equiv1-2\sin^2\theta_{ex}$ and quantify the present 
experimental status for $|\epsilon|<0.3$. We show that  both probabilities 
and observables depend on $\epsilon$ quadratically when effects are due 
to vacuum oscillations and they depend on $\epsilon$ linearly if matter 
effects dominate. The most important information on 
$\nu_e$--mixing comes from solar neutrino experiments. We find that 
the global analysis of solar neutrino data allows maximal 
mixing with confidence level better than 99\%
for $10^{-8}$ eV$^2\lsim\Delta m^2\lsim2\times10^{-7}$ eV$^2$. 
In the mass ranges $\Delta m^2\gsim 1.5\times10^{-5}$ eV$^2$ 
and $4\times10^{-10}$ eV$^2\lsim\Delta m^2\lsim2\times10^{-7}$
eV$^2$ the full interval $|\epsilon|<0.3$ is allowed within 
$\sim$ 4$\sigma$ (99.995 \% CL). We suggest ways to measure 
$\epsilon$ in future 
experiments. The observable that is most sensitive to $\epsilon$ 
is the rate [NC]/[CC] in combination with the Day--Night 
asymmetry in the SNO detector. With theoretical and statistical 
uncertainties, the expected accuracy after 5 years is  
$\Delta \epsilon\sim 0.07$. We also discuss the effects of 
maximal and near--maximal $\nu_e$-mixing in atmospheric 
neutrinos, supernova neutrinos, and neutrinoless double beta 
decay. 
\end{abstract}

\end{titlepage}


\section{Introduction}
\label{sec:introduction}
The data from both atmospheric and solar neutrino experiments  
have given a rather convincing evidence for non-zero 
neutrino masses 
and mixing. Furthermore, some features of the neutrino flavour
parameters are surprising and quite different from the quark 
flavour parameters. In particular, one or two of the three mixing 
angles in the MNS (Maki-Nakagawa-Sakata) mixing matrix for leptons
\cite{MNS}\ are large. Specifically, the simplest interpretation of 
the atmospheric neutrino measurements gives \cite{SKatm,FGV}
\begin{equation}
\sin^22\theta_{\mu\tau} \sim 0.8-1,\, \;\;\;\;\;
\Delta m^2_{32}\sim(2-5)\times 10^{-3}\,  \mbox{\rm eV}^2. 
\label{ANdata}  
\end{equation}
There exist several solutions of the solar neutrino problem
involving oscillations of electron neutrinos into some combination,
$\nu_x$, of $\mu-$ and  $\tau-$neutrinos with large mixing angle  
\cite{gonzalez,GGPG,three,four,lma,snoshow,dego} with parameters in 
the range  
\begin{equation}
\sin^22\theta_{ex} \sim 0.7-1,\, 
\;\;\;\Delta m^2_{12}\sim(0.2 - 4)\times 10^{-4}\,  \mbox{\rm eV}^2
\;\;\mbox{or} \;\;(0.05-20)\times 10^{-8}\,  \mbox{\rm eV}^2.
\label{SNdata}  
\end{equation}
 
The mixing angles involved in the explanation of
the solar and atmospheric neutrino data are not 
just order one. They are actually near--maximal, that is, 
$\sin^22\theta$  close to  $1$. If indeed one of the mixing
angles is near--maximal, it would provide a strong
support to the idea that the corresponding neutrinos are
in a pseudo-Dirac state. Such a scenario would have very
interesting implications for theoretical flavour models.
These implications have been recently studied in Ref.~\cite{nirpd}. 
A precise knowledge of the mixing and, in particular 
lower and/or upper bounds on small deviations from maximal mixing
provides an excellent probe of the related flavour physics.
It is the purpose of this work to study the experimental ways
in which the region of near--maximal mixing can be probed.

It is important to notice that there exists also a viable solution of 
the solar neutrino problem that does not involve large mixing, the so called
small mixing angle solution (SMA). Clearly, identification of the SMA 
solution will immediately exclude the possibility discussed in this 
paper. Also discovery of the sterile neutrinos will require a change 
of the whole picture.
Here we consider only three light active neutrinos and take the
$3\times3$ MNS matrix to be unitary.

Our main interest lies in the study of near--maximal mixing
involving $\nu_e$. We define a small parameter $\epsilon$ 
which describes the deviation from maximal mixing as: 
\begin{equation}
\sin^2\theta_{ex}\equiv{1\over2}(1-\epsilon),\ \ \ |\epsilon|\ll1. 
\label{defeps}  
\end{equation}
Our convention is such that $\Delta m^2_{21}\equiv m_2^2-m_1^2>0$. Then 
$\epsilon>0$ ($<0$) corresponds to the case that the lighter neutrino
mass eigenstate, $\nu_1$, has a larger (smaller) component of $\nu_e$
and the heavier one, $\nu_2$, has a larger (smaller) component
of $\nu_x$. 

Which value of deviation from maximal mixing is expected? 
In the case of theoretical models where the pseudo-Dirac structure is 
naturally induced, one expects that the deviation from maximal mixing 
is suppressed by a small parameter that is related to an approximate 
horizontal symmetry. If the same symmetry is also responsible for the 
smallness and hierarchy of the quark sector parameters, then it is 
quite plausible that $|\epsilon|\leq{\cal O}(\lambda)$, where 
$\lambda=0.22$ is the Cabibbo angle in the Wolfenstein parametrization.  
In various flavor models, the deviation from maximal mixing is related
to other physical parameters. For example, in a large class of models,
we have \cite{grni}\ 
\begin{equation}
|\epsilon|\gsim 2{m_e}/{m_{\mu}} \approx 0.01. 
\label{epslept}
\end{equation}
The $U(1)\times U(1)$ models described in refs. \cite{GNS,nirpd}\ give 
$|\epsilon|\sim(\Delta m^2_{12}/\Delta m^2_{23})^{1/3}$, while
models of $L_e-L_\mu-L_\tau$ symmetry \cite{baha,nirpd}\ give
$|\epsilon|\sim\Delta m^2_{12}/\Delta m^2_{23}$.
Quite generally we have 
$|\epsilon|\gsim\Delta m^2_{12}/\Delta m^2_{23}$,
which could be more restrictive than Eq.~(\ref{epslept})
if the solution to the solar neutrino problem lies at the
upper end of the $\Delta m^2_{12}$ range given in Eq.~(\ref{SNdata}).
In a large class of models we also have $|\epsilon|$ of the same
order of magnitude as $|U_{e3}|$, the mixing of $\nu_e$ in the 
third mass eigenstate, which can again be tested in the future 
\cite{nirpd}.

Large values of $\epsilon$, $|\epsilon|\gsim0.3$, will testify against 
at least the simplest versions of these theories. Therefore 
we consider both positive and negative values of $\epsilon$ in the range 
\begin{equation}
|\epsilon|\lsim0.3
\label{epsrange}
\end{equation}
which corresponds to $\sin^2 2\theta_{ex}\gsim0.91$. As concerns the 
mass-squared difference we cover the whole range below the reactor
bound:
\begin{equation}
10^{-11}\ \mbox{\rm eV}^2\lsim\Delta m^2_{21}\lsim10^{-3}\ \mbox{\rm eV}^2.
\label{dmsrange}
\end{equation}

Most of our discussion takes place in an effective two neutrino
generation framework. This is justified if $U_{e3}$ is zero or 
sufficiently small. We quantify this condition and consider the effect 
of a non-zero $U_{e3}$ in the Sect.~\ref{sec:threenus}.
We find there that, for matter oscillations, the leading corrections to 
the case of maximal mixing are of ${\cal O}(\epsilon,|U_{e3}|^2)$,
so that a reduction to a two neutrino analysis is justified for
$|U_{e3}|^2\ll\epsilon$. For vacuum oscillations, the corrections are
of ${\cal O}(\epsilon^2,|U_{e3}|^2)$, 
and a two neutrino analysis is valid for $|U_{e3}|\ll\epsilon$.

On the experimental side, we note that if 
$\Delta m^2_{{\rm atm}}>2\times10^{-3}\ \mbox{\rm eV}^2$, 
then the CHOOZ experiment limit~\cite{CHOOZ}\
requires $|U_{e3}|^2 \leq 0.05$ \cite{BWW,three}. 
For such values of $|U_{e3}|^2$, a two generation analysis of matter
(vacuum) effects is always valid for $|\epsilon| > 0.1 (0.3)$. On the 
theoretical side, in a large class of flavour models, $|U_{e3}|\sim
|\epsilon|$ \cite{nirpd}. In such a framework, a two generation analysis 
of matter effects is a good approximation while vacuum oscillations 
should be considered in the three generation framework.

In the limit $|U_{e3}|=0$ (which reduces the problem to a two neutrino 
framework), the deviation from maximal mixing can be determined as 
in (\ref{defeps}). In addition the mixing of $\nu_{\mu}$ and 
$\nu_{\tau}$ can also  be maximal, as favored by the atmospheric 
neutrino data. In this case we have the mixing structure
\begin{equation}
|U_{e3}|^2 = 0, \ \ \ \ 
|U_{e1}|^2=|U_{e2}|^2 =  \frac{1}{2} ,\ \ \ \  
|U_{\mu 3}|^2 = |U_{\tau 3}|^2 = \frac{1}{2} , 
\label{maxth3}  
\end{equation}  
which corresponds to the so called bi-maximal mixing scheme \cite{BPWW}. 
The analysis for the solar neutrino phenomenology is independent of 
$U_{\mu 3}$ and therefore the results discussed in Secs.~\ref{sec:status}, 
\ref{sec:predic}, \ref{sec:future}, \ref{subsec:supernova} 
and~\ref{subsec:double} apply also to the bi-maximal mixing case. Only in 
Sec.~\ref{subsec:atmos} where we discuss atmospheric neutrinos, does 
$U_{\mu3}$ play a role, and there we take it to be large and possibly 
maximal.

The plan of this paper is as follows. In Sec.~\ref{sec:physics} 
we give some basic physics considerations and useful expressions for 
the survival probability and for various observables.  
In Sec.~\ref{sec:status}  we describe the present
status of maximal mixing from solar neutrino experiments. 
The results of a global fit to all available solar neutrino 
data are given in Sec.~\ref{subsec:globalf}. The dependence of these 
results on various aspects of the analysis is  described in 
Sec.~\ref{subsec:dependence}. 
In Sec.~\ref{sec:predic} we study the predictions that follow from 
near--maximal mixing for individual, existing measurements: 
total rates, Argon production rate,  Germanium production rate, 
the Day--Night asymmetry in elastic scattering events, 
the zenith angle distribution of elastic scattering events, and 
the shape of the recoil energy spectrum.
In Sec.~\ref{sec:future} we suggest tests of maximal mixing from
future experiments. We describe the conditions for having unambiguous
tests in Sec.~\ref{subsec:requirements}. Then we examine individual 
experiments: GNO and Super--Kamiokande, SNO, 
Borexino and low energy solar neutrino experiments. 
The effect of extending to three neutrino
scenario is commented in Sec.~\ref{sec:threenus}.
In Sec.~\ref{sec:other} we discuss the effect of maximal mixing in 
atmospheric neutrinos, supernova neutrinos and neutrinoless double beta 
decay.  We present our conclusions in Sec.~\ref{sec:conclusion}.

\section{Physics at near--maximal mixing}
\label{sec:physics} 

\subsection{The survival probability for solar neutrinos}
\label{subsec:prob}
In this subsection we present general  expressions for the survival
probability of solar electron  neutrino in a two generation framework
valid
in the full range of $\Delta m^2$ which we use in our numerical
calculations. 
 
The survival amplitude for a solar $\nu_e$ neutrino of
energy $E$ at a detector in the Earth can be written as:
\begin{equation}
A_{ee}=\sum_{i=1}^2 A^S_{e\,i}\,A^E_{i\,e}\,\exp[-im_i^2 (L-r)/2E]~. 
\end{equation}
Here $A^S_{e\,i}$ is the amplitude  of the transition  
$\nu_e \rightarrow \nu_i$ ($\nu_i$ is the $i$-mass eigenstate) 
from the production point to the Sun surface,  $A^E_{i\,e}$ is the 
amplitude of the transition  $\nu_i \rightarrow \nu_e$ 
from the Earth surface to the detector, and the propagation in vacuum 
from the Sun to the surface of the  Earth is given by the exponential. 
$L$ is the distance between the center of the Sun and the surface of 
the Earth, and $r$ is the distance between the neutrino production 
point and the surface of the Sun.  
The corresponding survival probability $P_{ee}$ is then given by:
\begin{eqnarray} \label{Pee}
P_{ee}&=&P_1P_{1e}+P_2P_{2e}+2\sqrt{P_1P_2P_{1e}P_{2e}}\cos\xi\nonumber\\
&=&P_1+(1-2P_1)P_{2e}+2\sqrt{P_1(1-P_1)P_{2e}(1-P_{2e})}\cos\xi.
\end{eqnarray}
Here $P_i\equiv |A^S_{e\,i}|^2$ is the probability that the solar 
neutrinos reach the surface of the Sun  as $|\nu_i\rangle$ and
we use $P_1+P_2=1$, while $P_{ie} \equiv  |A^E_{i\,e}|^2$ is the 
probability of $\nu_i$ arriving at the surface of the Earth to be 
detected as a $\nu_e$, and we use $P_{1e}+P_{2e}=1$. The phase $\xi$ is 
given by 
\begin{equation} 
\xi=\frac{\Delta m^2 (L-r)}{2E}+\delta\, ,
\end{equation}
where $\delta$ contains the phases due to propagation in the
Sun and in the Earth and can be safely neglected since it is always
much smaller than the preceeding term \cite{panta,threev}.

From  Eq.~(\ref{Pee}) one can recover more familiar expressions for 
$P_{ee}$:  

(1) For $\Delta m^2/E\lesssim 5 \times 10^{-17}$ eV, the matter effect 
supresses flavour transitions in both the Sun and the Earth. 
Consequently, the probabilities $P_1$ and $P_{2e}$ are simply
the projections of the $\nu_e$ state onto the mass eigenstates:
$P_1 = \cos^2\theta$,    $P_{2e} = \sin^2 \theta$. 
In this case we are left with the 
standard vacuum oscillation formula:
\begin{equation}
P_{ee}^{\rm vac}=1-\sin^2 2\theta \sin^2(\Delta m^2 (L-r)/4E)
\label{pvac}
\end{equation}
which describes the oscillations on the way from the surface of the Sun to 
the surface of the Earth. The probability is symmetric under  
$\theta \leftrightarrow \frac{\pi}{2}-\theta$.

(2) For $\Delta m^2/E\gtrsim 10^{-14}$ eV, the last term in Eq.~(\ref{Pee})
vanishes and we recover the incoherent MSW survival probability.
For $\Delta m^2/E\sim10^{-14}-10^{-12}$ eV$^2$, this term is
zero because $\nu_e$ adiabatically converts to $\nu_2$ and $P_1=0$.
For $\Delta m^2/E\gsim10^{-12}$ eV$^2$, both $P_1$ and $P_2$ are nonzero 
and the term vanishes due to averaging of $\cos\xi$.

(3) In the intermediate range, 
$5 \times 10^{-17} \lesssim  \Delta m^2/E  \lesssim 10^{-14}$ eV, 
adiabaticity is violated and the $\cos\xi$ coherent term should be 
taken into account. The result is similar to vacuum oscillations 
but with small matter corrections. We define this case as  quasi-vacuum 
oscillations \cite{panta,threev,fried}.

The results presented in the following sections have been obtained using
the general expression for the survival probability in Eq.~(\ref{Pee})
with $P_1$ and $P_{2e}$ found by numerically solving the evolution
equation 
in the Sun and the Earth matter. For $P_i$ we use the electron number 
density of 
BP2000 model \cite{BP00}. For $P_{ie}$ we integrate numerically the evolution 
equation in the Earth matter using the Earth density profile given in the 
Preliminary Reference Earth Model (PREM) \cite{PREM}. 

\subsection{The mixing angle in matter}

While, as explained above, our results are obtained by a numerical
calculation, it is useful to find  approximate analytical expressions 
for the neutrino survival probability and for various observables.
This is done in this and in the following subsections. The analytical 
expressions help us to qualitatively understand the behaviour of the 
different observables, particularly for the case of 
near--maximal mixing. 

The probabilities and observables depend on the mixing angle in matter 
$\theta_m$ via $\cos 2\theta_m$ which enters the probability  
of the adiabatic conversion and via $\sin^22\theta_m$ which determines,  
{\it e.g.},  the depth of oscillations in a uniform medium.  

We can write $\cos2\theta_m$ in terms of the neutrino 
oscillation parameters and the electron density in medium as:
\begin{equation}
\cos2\theta_m = {-1 + \eta \cos2\theta\over(
1-2\eta \cos2\theta+\eta^2)^{1/2}}.
\label{tSnusun}
\end{equation}
Here
\begin{equation}
\eta\equiv\frac{l_0}{l_\nu}
=0.66\left({\Delta m^2/E\over10^{-15}\ \mbox{\rm eV}}\right)
\left({10^{-2}\ \mbox{\rm g\ cm}^{-3}\over\rho Y_e}\right) 
\label{defeta}
\end{equation}
is the ratio between the refraction length, $l_0$, and the neutrino 
oscillation length in vacuum, $l_\nu$:
\begin{eqnarray}
l_0\equiv\frac{2\pi m_N}{\sqrt2 G_F\rho Y_e},  &
& \ \ \ \ l_\nu\equiv\frac{4\pi E}{\Delta m^2}~. 
\end{eqnarray}
Here $\rho$ is the matter density and $Y_e$ is the number of electrons 
per nucleon. 

Around maximal mixing $\cos2\theta_m$ can be expanded as 
\begin{equation}
\cos2\theta_m = \ -{1\over\sqrt{1+\eta^2}}\left(1-
{\eta^3\over1+\eta^2}\epsilon\right). 
\label{ctseps}
\end{equation}
In the limit of weak matter effects, $\eta \gg 1$, and in the matter
dominance case, $\eta \ll 1$, we get:
\begin{equation}
\cos2\theta_m = \left\{
\begin{array}{ll}
\epsilon\ \ \ \ \ &\eta \gg 1\\
- 1\ \ \ \ \ & \eta \ll 1 
\end{array}
\right. .
\label{ctseta}
\end{equation}
The dependence of $\cos2\theta_m$ on $\epsilon$ is smooth.
It is stronger for $\eta\gg1$ and highly suppressed for $\eta\ll1$. 
For precisely maximal mixing we have $\cos2\theta_m=-1/\sqrt{1+\eta^2}$, 
which decreases from zero in vacuum to $-1$ in the matter dominance 
case. 

The  expression for $\sin^2 2\theta_m$ 
for near maximal mixing is given by 
\begin{equation}
\sin^2 2\theta_m = \ {{\eta^2} \over {1+\eta^2}} 
\left(1 + {2 \eta \over1+\eta^2}\epsilon\right),
\label{stseps} 
\end{equation}
which leads to
\begin{equation}
\sin^2 2\theta_m = 
\left\{
\begin{array}{ll}
1 +  \frac{2}{\eta} \epsilon \ \ \ \ & \eta \gg 1  \\ 
\eta^2 (1 + 2 \eta  \epsilon)\ \ \ \ \  &\eta \ll 1 
\end{array}
\right.  .
\label{stseta}
\end{equation}
In both cases the $\epsilon$  corrections to the value at maximal mixing
are strongly suppressed.  In vacuum, the correction is quadratic in 
$\epsilon$: $\sin^2 2\theta = 1 - \epsilon^2$. 

\subsection{Survival probability}
\label{subsec:analprob}

We first consider the survival probability for electron neutrinos 
without the regeneration effect in the Earth. It describes the $\nu_e$ 
flux arriving at the Earth during the day. In daytime, 
$P_{2e}=\sin^2\theta$. Consequently, Eq.~(\ref{Pee}) gives,  
in the region where the oscillating term in Eq.~(\ref{Pee}) is absent,
\begin{equation}
P_D\ =\  \frac{1}{2} + \epsilon \left(P_1 -\frac{1}{2}\right).  
\label{Psubd}
\end{equation}
The neutrino evolution in the Sun described by the probability $P_1$ 
can be approximated by the well known expression
\begin{equation}
P_1 = \frac{1}{2} + \left(\frac{1}{2} - P_c \right)\cos2\theta_S.  
\label{Ponead}
\end{equation}
Here $\theta_S$ is the matter mixing angle at the production point:
\begin{equation}
\cos 2\theta_S \equiv \cos 2 \theta_m(\eta_S), ~~~~ \eta_S \equiv 
\eta(\rho_S Y_{eS}),   
\label{cos}
\end{equation}
where $\rho_S$ and $Y_{eS}$ are, respectively, the matter density
and the number of electrons per nucleon  at the production 
point. Eq.~(\ref{Ponead}) is assumed to be averaged over 
the production region in the Sun. $P_c$ is the jump  probability which 
describes the violation of adiabaticity. For an exponential density 
profile it takes the following form \cite{Petc,KrPe}: 
\begin{equation}
P_c = {e^{-\gamma\sin^2\theta}-e^{-\gamma}\over1-e^{-\gamma}} =  
{e^{-(\gamma/2)(1-\epsilon)}-e^{-\gamma}\over1-e^{-\gamma}}, 
\label{finPc}
\end{equation}
where $\gamma$ is the ratio of the density scale height $l_\rho$ and 
the neutrino oscillation length:
\begin{eqnarray}
\gamma\equiv\frac{4\pi^2 l_\rho}{l_\nu}=
1.05\left({\Delta m^2/E\over10^{-15}\ \mbox{\rm eV}}\right)
\left({l_\rho\over r_0}\right),&  & ~~~~
l_\rho\equiv{\rho\over d\rho/dr}.
\label{defgam}
\end{eqnarray}
The length scale $r_0=R_\odot/10.54$ is related to the exponential
approximation to the solar density profile, $\rho=\rho_0\exp(-r/r_0)$.

Notice that, originally, Eq.~(\ref{finPc}) was derived for a 
mixing angle $\theta<\frac{\pi}{4}$ where resonant enhancement is 
possible. However both Eq.~(\ref{tSnusun}) and Eq.~(\ref{finPc}) 
can be analytically continued into the second octant, 
$\theta > \pi/4$, and used to 
compute the corresponding survival probability for $\epsilon<0$ 
\cite{dego,GGPG,three,four}. 

Inserting $P_1$ of Eq.~(\ref{Ponead}) into Eq.~(\ref{Psubd}), we get 
\begin{equation}
P_D=\frac{1}{2} + \epsilon \left(\frac{1}{2}-P_c\right) \cos2\theta_S.
\label{psubd}
\end{equation}

\begin{figure}[!t]
\centerline{\psfig{figure=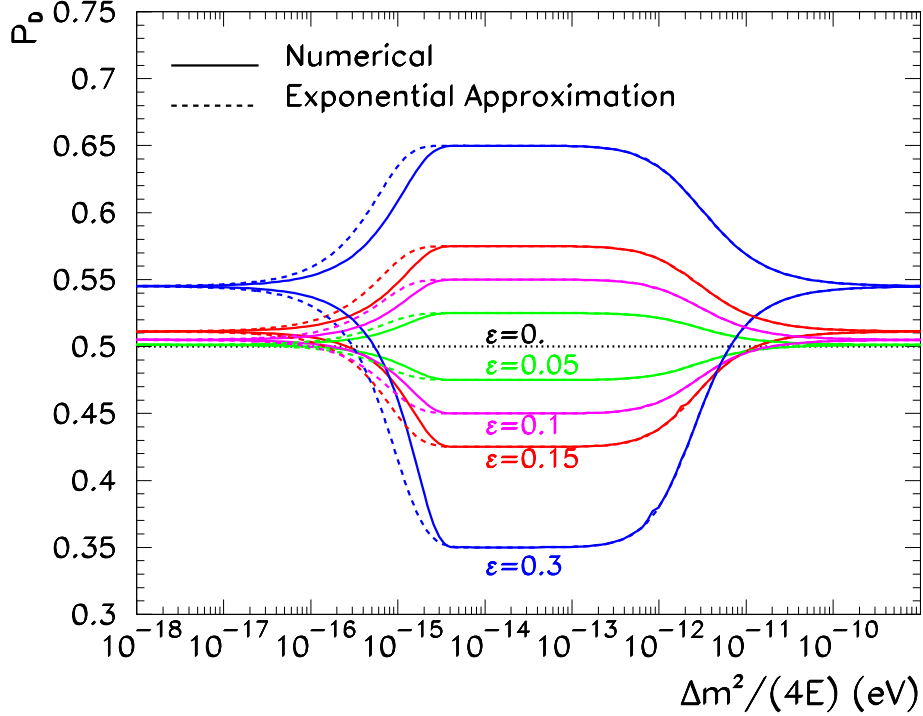,width=5.0in,angle=0}}
\tightenlines
\caption[]{\small The dependence of the averaged $\nu_e$ survival
probability 
inside the sun on $\Delta m^2/4E$ for different values of 
$\epsilon$ (numbers at the curves). The upper curves describe the 
corresponding negative values of $\epsilon$. Solid lines show results of
numerical calculations. Dashed lines correspond to analytical formulae 
with exponential approximation for the solar density profile. 
\label{fig:sun}}
\end{figure}

Let us study the properties of $P_D$. In Fig.~\ref{fig:sun} we plot 
$P_D$  as a function of $\Delta m^2/4E$ for different values of the 
deviation from maximal mixing. We show in the figure the
results obtained by the numerical calculation as well as from the
corresponding analytical approximation (\ref{finPc}) for 
exponential density profile. We learn the following points from
Eq.~(\ref{psubd}) and Fig.~\ref{fig:sun}:

1) For maximal mixing ($\epsilon=0$), $P_D = 1/2$ independently of 
the adiabaticity violation (encoded in $P_c$), matter effects, etc.. 

2) For near-maximal mixing ($\epsilon \neq 0$),  solar matter effects  
lead to an energy dependent probability in the MSW region  
$4 \times 10^{-16} \lesssim  \Delta m^2/E  \lesssim 10^{-10}$ eV. 
 
3) For $\Delta m^2/E\gg10^{-15}$ eV, the adiabaticity condition, 
$\gamma\gg1$, is fulfilled (see Eq.~(\ref{defgam})). Consequently 
$P_c=0$ and Eq.~(\ref{psubd}) gives 
\begin{equation}
P_D=\frac{1}{2}\left(1+\epsilon\cos2\theta_S\right)\ \ \ \ \ \ 
(\Delta m^2/E\gg10^{-15}\ \mbox{\rm eV}).
\label{pdepsH}
\end{equation}
From Fig.~\ref{fig:sun}\ we see that the analytical expression gives a 
good description of the propagation in the Sun for this region. 

4) For $\eta_S\gg1$ (weak matter effect), we have $\cos2\theta_S=\epsilon$ 
(see Eq.~(\ref{ctseta})) and the probability (\ref{pdepsH}) reduces to
the vacuum oscillation probability:
\begin{equation}
P_D=\frac{1}{2}\left(1+\epsilon^2\right)\ \ \ \ \ \ 
(\Delta m^2/E\gg10^{-11}\ \mbox{\rm eV}). 
\label{pdepsL}
\end{equation}

5) For  $\eta_S \ll 1$ (the resonance layer is far enough from the 
production point which is close to the center of the sun), 
we have $\cos2\theta_S = -1$ and 
\begin{equation}
P_D=\frac{1}{2}\left(1-\epsilon\right)\ \ \ \ \ \ 
(10^{-15}\ \mbox{\rm eV}\ll\Delta m^2/E\ll10^{-11}\ \mbox{\rm eV}).
\label{pdepsI}
\end{equation}
This is the region of strong adiabatic conversion, when $\nu_e$
produced in the center of the Sun practically coincides with the matter 
eigenstate $\nu_{2m}$ at the production point and during its propagation
inside the Sun. Therefore, it emerges from the Sun and
reaches Earth as $\nu_{2}$.

6) For $\Delta m^2/E < 10^{-15}$ eV, adiabaticity is violated.   
We see from Fig.~\ref{fig:sun}\ that the analytical results obtained 
for an exponential density profile differ  from the results of 
the numerical calculations. In particular, the analytical result shows  
a ``slower'' transition to the vacuum oscillation regime or, in other
words, it overestimates the size of the matter effects in the 
quasi--vacuum oscillation region. The same value of the survival
probability appears for about two times smaller $\Delta m^2/E$. 
Similar conclusions have been drawn in refs. \cite{threev,fried}\ .

7) For $\epsilon\neq 0$, the effects of the adiabatic edge situated at
$\Delta m^2/E=(10^{-12}-10^{-10})$ eV, and of the non-adiabatic edge
situated at $\Delta m^2/E=(10^{-16}-3\times10^{-15})$ 
eV, become important.

8) For  $\Delta m^2/E\ll10^{-17}$ eV, as we noted in 
Sec. \ref{subsec:prob}, the effect is reduced to the vacuum oscillation 
between the surfaces of the Sun and the Earth and the average 
survival probability, $P_D=1-\frac{1}{2}\sin^22\theta$, 
(shown in Fig.~\ref{fig:sun})  coincides with 
that in Eq.~(\ref{pdepsL}). However, in this region  averaging 
of   oscillations does not occur and for the survival probability we 
should use 
\begin{equation}
P_{ee}
\approx 1-(1 - \epsilon^2)\sin^2\frac{\phi}{2}~, 
\label{avprob2}
\end{equation}
where $\phi = \Delta m^2 (L - r)/4E$ is the oscillation phase which  does
not depend on
$\epsilon$. In principle matter effects strongly suppress the 
oscillations inside the Sun and the Earth. However, the modification of 
the observables is small, since the size of the Sun  (and the Earth) is 
much smaller than the oscillation length in vacuum. 
The quadratic dependence of the probability on $\epsilon$ is 
again restored.  Moreover, time variations of signals and the distortion
of the energy spectrum  originate from the dependence of the phase on 
$L$ and $E$. Therefore according to (\ref{avprob2}), both 
the variations and the distortion are 
proportional to $(1 - \epsilon^2)$. That is, the 
dependence of all observables on $\epsilon^2$ near maximal mixing 
is very weak.

According to Eq.~(\ref{pdepsI}) and Fig.~\ref{fig:sun}, in the 
MSW region (more precisely at the bottom of the suppression pit), the
survival probability depends on $\epsilon$ linearly. For $\Delta m^2/E$ 
below and above the MSW region the probability converges to the vacuum 
oscillation probability. The deviation of the latter from the 
probability at maximal mixing depends on $\epsilon$ quadratically and
the dependence on the sign of  $\epsilon$ disappears. From this we 
infer that the sensitivity of $P_D$ and consequently of observables to 
$\epsilon$ is much higher in the MSW region. 

Notice that there are two transition regions (between MSW and the vacuum
oscillation regions) where the effects are mainly due to vacuum
oscillations with small matter corrections. 
We call these regions `Quasi-Vacuum Oscillation regions': we denote 
by QVO$_{\rm L}$ (QVO$_{\rm S}$) the one with $\Delta m^2$ larger 
(smaller) than in the  MSW region. In the QVO$_{\rm L}$ and 
QVO$_{\rm S}$ regions, the linear dependence of the probability and the 
observables transforms into quadratic dependence.

\subsection{Regeneration effects in the Earth}
\label{subsec:analreg}

For a neutrino arriving at night time,  Earth matter effects should be 
taken into account. To gain a qualitative understanding of the Earth 
effects, we make the crude approximation of uniform density, such that 
the mixing angle in the Earth is constant, $\theta_E$, along the 
neutrino trajectory. In this case, the neutrino propagation has an 
oscillatory character. We assume that the oscillations are 
averaged out, and therefore 
\begin{eqnarray}\label{Ptwoe}
P_{2e}(\mbox{\rm vacuum})&=&\sin^2\theta,\nonumber\\
P_{2e}(\mbox{\rm uniform\ density})&=&{1\over2}\left[
\sin^2\theta+\sin^2(2\theta_E-\theta)\right].
\end{eqnarray}
It is convenient to introduce  a regeneration factor which describes the
Earth matter effect:
\begin{equation}
\freg \equiv P_{2e}({\mbox{\rm matter}})-P_{2e}({\mbox{\rm vacuum}})
=\ {1\over2}\sin2\theta_E\sin(2\theta_E-2\theta).
\label{regen}
\end{equation}
Denoting by $\eta_E$ the parameter $\eta$ in the Earth: 
\begin{equation}
\eta_E \equiv  \eta ( \rho_E Y_{eE} ), 
\label{etaE}
\end{equation} 
where  $\eta$ is defined in Eq.~(\ref{defeta}), we find from 
Eq.~(\ref{regen}):
\begin{equation}
\freg={\eta_E\sin^22\theta\over2(1-2\eta_E\cos2\theta+\eta_E^2)}.
\label{fregear}
\end{equation}
Note that $\freg$ is always (for any value $\Delta m^2/E$ and $\epsilon$) 
positive, {\it i.e.} the matter effect of
Earth always {\it enhances} the survival probability $P_{ee}$.

The parameter $\freg$ can be expanded around $\epsilon=0$:
\begin{equation}
\freg=\ {\eta_E\over2(1+\eta_E^2)}\left(
1+{2\eta_E\over1+\eta_E^2}\epsilon\right). 
\label{freeps}
\end{equation}
The regeneration factor has a maximum at $\eta_E \approx 1$ which 
corresponds to  $\Delta m^2/E \sim (2 - 3)\times 10^{-13}$ eV. 
This determines the {\it strong regeneration region} in the 
$\Delta m^2/E$ scale. It is situated in the middle of the the solar MSW 
region.  Strong regeneration effects are already excluded by the 
SuperKamiokande result on Day-Night asymmetry \cite{superk,suzuki}. 
Consequently, the strong regeneration region separates two (allowed) 
parts within the MSW region in which the regeneration effects are small: 

1. HIGH  $\Delta m^2$ region,  where $\eta_E \gg 1$ and 
$\freg\sim 1/\eta_E$; 

2. LOW $\Delta m^2$ region, where $\eta_E \ll 1$ and
$\freg\sim \eta_E/2$. 

We will quantify the borders of these regions in Sec. III.  
In both cases the regeneration effect is suppressed by a small parameter
and it disappears when moving away from the strong regeneration region. 

Let us stress that the SuperKamiokande limit on regeneration effects 
holds for the energy range $E = 5 - 15$ MeV to which this experiment is 
sensitive. This corresponds to $\Delta m^2=(2-3)\times10^{-6}$ eV$^2$. 
However, strong regeneration effects are not excluded for other 
energies, in particular, for low energy neutrinos. In this case 
$\eta_E$ can be of the order one and $\freg$ at its maximum.

To first order in $\epsilon$, we obtain from Eq.~(\ref{freeps}): 
\begin{equation}
\freg = \left\{
\begin{array}{ll}
\frac{1}{2\eta_E}
\left(1 + \frac{2}{\eta_E} \epsilon  \right) \ \ \ \ \ &({\rm HIGH})\\
\frac{\eta_E}{2} (1 + 2 \eta_E \epsilon)   \ \ \ \ \ &(\rm LOW) 
\end{array}
\right.  .  
\label{freglim}
\end{equation}
In both the HIGH and LOW regions the dependence of the regeneration 
factor on $\epsilon$ is further suppressed by the small parameter 
min$\{2/\eta_E, 2 \eta_E \}$.

The probability $P_{ee}$ at night time, $P_N$, can be written (for the 
region where the oscillating term in Eq.~(\ref{Pee}) is absent) 
as follows:
\begin{equation}
P_N\ =\ P_1 + (1-2P_1)\left(\sin^2 \theta + \freg\right) = 
\frac{1}{2}+\frac{1}{2}(1-2P_c)\cos2\theta_S(\cos2\theta-2\freg).  
\label{Pnig}
\end{equation}
The average daily survival probability is given for $\eta_S \ll 1$ 
($\cos2\theta_S \approx -1$) by
\begin{equation}
\bar P\equiv{1\over2}(P_D+P_N)
={1\over2}[1+(1-2P_1)(\freg-\cos2\theta)].
\label{barP}
\end{equation}
The Day--Night asymmetry is given by
\begin{equation}
A_{\rm N-D}\equiv{P_N-P_D\over\bar P}
={2\freg\over1/(1-2P_1)-\cos2\theta+\freg}.
\label{defAND}
\end{equation}

Let us now consider the dependence of $A_{N-D}$ and $\bar P$ on
$\epsilon$ in the HIGH and in the LOW regions keeping the lowest 
order terms in $\epsilon$ and in min$\{\eta_E, 1/\eta_E\}$. 

1. In the HIGH region, the adiabaticity condition is satisfied and 
we can safely put $P_c=0$. Consequently,
$P_1=\cos^2\theta_S$. Then, Eq.~(\ref{barP}) simplifies:
\begin{equation}
\bar P={1\over2}[1-\cos2\theta_S(\freg-\cos2\theta)].
\label{Plma}
\end{equation}
With near-maximal mixing, we obtain:
\begin{equation}
\bar P\ =\ {1\over2}\left(1- 
\frac{\epsilon - 1/2\eta_E}{\sqrt{1+\eta_S^2}}\right)
+{\cal O}\left({1\over\eta_E\eta_S},{\epsilon\over\eta_E}\right).
\label{Plmactm} 
\end{equation}
For  $\eta_S^2 \ll 1$ (small $\Delta m^2$) we find: 
\begin{equation}
\bar P\approx \frac{1}{2}\left(1-\epsilon+\frac{1}{2\eta_E}+ 
\frac{\epsilon}{\eta_E^2}\right) ~, 
\label{adiabprob1}
\end{equation}
where the last term is a small correction which comes from the 
regeneration factor. For the Day--Night asymmetry we obtain:
\begin{equation}
A_{\rm N-D}\ =\ 
{1\over\eta_E\sqrt{1+\eta_S^2}}\left[1+{
(\sqrt{1+\eta_S^2}-\eta_S^3)\epsilon\over1+\eta_S^2}
+{\cal O}\left({1\over\eta_E^2},{\epsilon\over\eta_E}\right)\right].
\label{Almactm} 
\end{equation}
Notice that the $\epsilon$-dependent effect changes sign between large
and small $\eta_S$. For $\eta_S^2 \ll 1$ we get: 
\begin{equation} 
A_{\rm N-D}
\simeq \frac{1}{\eta_E}(1+\epsilon)~.
\label{adnlma}
\end{equation}
The asymmetry increases linearly with $\epsilon$ for small
enough $\eta_S^2$. 

2. In the LOW region we can take  $\eta_S=0$ and, consequently,
$P_1=P_c$. Then, Eq.~(\ref{barP}) simplifies:
\begin{equation}   
\bar P = \frac{1}{2} \left[ 1 + (1 - 2P_c)(\freg- \epsilon)\right].   
\label{adiabprob2}
\end{equation}
In LOW  region we have $\gamma \gg 1$ (see Eq.~(\ref{defgam})) and,
therefore, $P_c\ll 1$ (see Eq.~(\ref{finPc})). In any case, the
$P_c$-dependent term in Eq.~(\ref{adiabprob2}) is suppressed by a small 
factor and can be neglected. We get: 
\begin{equation}
\bar P={1\over2}\left(1-\epsilon+{\eta_E\over2}\right).
\label{Plowctm}
\end{equation} 
For the Day-Night asymmetry we obtain taking $P_c \ll 1$:    
\begin{equation}
A_{\rm N-D} \approx {2\freg\over 1 + 2P_c - \epsilon  + \freg}, 
\label{defAND1} 
\end{equation} 
and in the limit $\eta_E \ll 1$: 
\begin{equation}
A_{{\rm N-D}}=\eta_E\left(1-2e^{-\gamma/2}-{\eta_E\over2}
+\epsilon\right).
\label{Alowctm} 
\end{equation}
A few comments are in order. The Day--Night asymmetry 
is suppressed by $\eta_E$. 
The $\epsilon$-dependent effect is one of several small
corrections to the leading result. For large parts of the LOW region
it is the leading correction, but for large $\Delta m^2/E$,
the subleading regeneration effect could be comparable while for
small $\Delta m^2/E$, the non-adiabatic correction could
give the main correction. 

As we mentioned before, a strong regeneration effect with 
$\eta \sim 1$ and   $\freg\sim 1 $  is not excluded for low energy
neutrinos. In particular in the LOW region strong regeneration 
can show up for the beryllium and pp-neutrino components of the
spectrum. In this case, the approximation $\eta_E\ll1$ does not work 
and one should use the complete expression for the regeneration factor 
and the asymmetry. 

As seen from Eqs.~(\ref{adiabprob1}), (\ref{Plowctm}), (\ref{adnlma}) 
and (\ref{Alowctm}), the dominant dependence on $\epsilon$ of both  
$\bar P$ and $A_{\rm N-D}$ arises from the dependence of the solar 
survival probability on $\epsilon$. The dependence which follows from 
the regeneration factor is further supressed by 
min$\{\eta_E,1/\eta_E\}$. (For $A_{\rm N-D}$ the $\epsilon$ dependence 
follows from the dependence on $\bar P$ in the denominator.)  

From Eqs.~(\ref{adnlma}) and~(\ref{Alowctm}) we find that the
Day-Night asymmetry strongly depends on $\Delta m^2$ due to the 
$\eta_E$ or $1/\eta_E$ factors. In the HIGH region $A_{\rm N-D} 
\propto \Delta m^2$  while in LOW region $A_{\rm N-D} \propto
1/\Delta m^2$. The dependence of the asymmetry on $\epsilon$ is much 
weaker: In both regions $A_{\rm N-D} \propto (1 + \epsilon)$. 
Consequently, the measurements of $A_{\rm N-D}$ are very sensitive to 
$\Delta m^2$ while the sensitivity to $\epsilon$ is substantially lower. 

Let us comment on the range of validity of the approximate treatment of 
the Earth effects \cite{dGFM}. In the HIGH region, for $E\sim$ MeV and 
$\Delta m^2\gg10^{-6} \mbox{\rm eV}^2$, the oscillation length 
$l_\nu$ is much shorter than the size of the Earth and neutrinos
undergo many oscillations inside the Earth. The constant electron number 
density approximation gives a good description of $A_{\rm N-D}$ which 
involves integration over the zenith angle. In the LOW region, for 
$\Delta m^2\ll 10^{-6} \mbox{\rm eV}^2$, the oscillation length is 
approximately equal to the refraction length $l_0$ and the latter is 
comparable to the size of the Earth (independently of $\Delta m^2$). 
The details of the $N_e$ profile do not play an important role, 
and the effect is determined by the average density along the neutrino
trajectory. 

Regeneration effect leads also to seasonal variations of signals 
\cite{dGFM,season,FLMP2}. These variations, however, are less sensitive to 
the oscillation parameters.

\subsection{Distortion of the energy spectrum}
\label{subsec:analspec}

The distortion of the solar neutrino energy spectrum can be characterized 
by the  distortion parameter defined as: 
\begin{equation}
s_{\nu} \equiv \frac{E}{\bar P}\frac{d \bar P}{dE}. 
\label{slope}
\end{equation}
Averaged over the appropriate energy interval this parameter  
is proportional to the shift of the first moment of the spectrum 
or to the slope parameter used in the literature. 

To understand the distortion of the spectrum (energy dependence of the 
averaged probability), we remind the reader that $\eta\propto1/E$
and use Eq.~(\ref{Plma}) for the HIGH and QVO$_{\rm L}$ regions
and Eq.~(\ref{adiabprob2}) for the LOW region.

1. The HIGH and QVO$_{\rm L}$ regions:
 
(i)  For large $\Delta m^2$, the effect of the adiabatic edge of the 
suppression pit which is encoded in the $\eta_S$ dependence is important:
\begin{equation}
s_{\nu} \approx  - \epsilon \frac{\eta_S^2}{(1 + \eta_S^2)^{3/2}}.   
\label{slopeHl}
\end{equation}
The distortion parameter is proportional to $\epsilon$.
The slope (shift of the first moment) is positive (negative) for  
$\epsilon<0$ ($\epsilon > 0$). The distortion decreases rapidly with  
$\eta_S$.  

(ii) For small $\Delta m^2$, the earth regeneration which is related to 
the $\eta_E$ dependence is important: 
\begin{equation}
s_{\nu} \approx  \frac{1}{2\eta_E} \left(1 + \epsilon 
- \frac{1}{2\eta_E} + \frac{4\epsilon}{\eta_E} \right). 
\label{slopeHs}
\end{equation}
The distortion decreases with increase of $\eta_E$ or 
$\Delta m^2$. The sensitivity to $\epsilon$ is much weaker than in the
previous case and it follows mainly from the dependence on the average 
probability in the denominator.  

2. The LOW region:

(i) For large $\Delta m^2$, the regeneration effect is important:
\begin{equation}
s_{\nu} \approx  \frac{\eta_E}{2} (1 + \epsilon). 
\label{slopeLh}
\end{equation}   
The slope increases with $\eta_E$. The dependence on $\epsilon$ is weak.

(ii) For small $\Delta m^2$, the effect of the nonadiabatic edge of the 
solar suppression pit gives the  dominant effect:
\begin{equation}
s_{\nu} =  2\epsilon E \frac{d P_c}{dE} 
\approx \epsilon \gamma e^{-\gamma/2}.   
\label{slopeLl}
\end{equation}
The distortion  is proportional to $\epsilon$. The slope is positive 
(negative) for $\epsilon > 0$ ($\epsilon < 0$). The effect is 
suppressed for relatively weak violation of the adiabaticity. 

As we have mentioned previously, for non-averaged vacuum oscillations 
$s_{\nu} \propto   (1 - \epsilon^2)$, that is, the dependence of
distortion on $\epsilon^2$ is very weak.

\subsection{Summary}
\label{subsec:analsum}

Let us summarize the results of our analytical studies. 
We have found simple analytical expressions for various 
observables (rates of events given by the survival probability, 
Day-Night asymmetry, distortion of the spectrum etc.) in 
terms of $\Delta m^2$ and the deviation from maximal mixing $\epsilon$. 
These approximate analytical expressions reproduce correctly the
functional dependence of the observables on $\Delta m^2$ and $\epsilon$ 
and allow us to understand all the features of the exact 
numerical calculations.

The effects and the dependence of observables on $\epsilon$ change with 
$\Delta m^2$. Accordingly, we define several regions of $\Delta m^2$ 
which have different physical pictures. As $\Delta m^2$ decreases from 
its upper (CHOOZ) bound, we have the following regions 
(we quantify borders of these regions in the next section):

\begin{itemize}

\item
quasi-vacuum oscillation region with large $\Delta m^2$ 
(QVO$_{{\rm L}}$),

\item
MSW region with high $\Delta m^2$  (HIGH),

\item
MSW region with low $\Delta m^2$  (LOW),

\item
quasi-vacuum oscillation region with small $\Delta m^2$
(QVO$_{{\rm S}}$),

\item
region of non-averaged vacuum oscillations (VO). 
The high $\Delta m^2$ part of the VO region will be called 
 VAC$_{\rm L}$.

\end{itemize}

As concerns the $\epsilon$ dependence of observables in these regions,
we find  two main conclusions:

1. Maximal mixing, $\epsilon = 0$, is not a special point 
as far as the  phenomenology is concerned (in contrast with theory). 
No divergencies or discontinuities appear in the dependence of
observables on $\epsilon$. The dependence  on $\epsilon$ 
is smooth and, in many cases, very weak. To mention a few examples: 
The day time survival probability is 1/2 at $\epsilon=0$.  
Earth regeneration effects, however, enhance the survival probability. 
In the MSW region the slope of the energy spectrum distortion is
proportional to $\epsilon$ in certain regions of $\Delta m^2$,
and consequently changes sign at $\epsilon=0$.  
The Day-Night asymmetry is proportional to $(1 + \epsilon)$. \\

2. The character of the $\epsilon$-dependence of observables is different 
in the vacuum oscillations and MSW regions. In the regions of vacuum 
oscillation all the effects depend on $\epsilon$ quadratically. 
More precisely, the $\epsilon$-dependent factors are of two types:
\begin{eqnarray}\label{avprob1}
1 + \epsilon^2&~~~~~~~~~~&({\rm averaged\ oscillations}),\nonumber\\
1 - \epsilon^2&~~~~~~~~~~&({\rm non-averaged\ oscillations}).
\end{eqnarray}
The dependence is symmetric 
with respect to interchanging $\epsilon \leftrightarrow - \epsilon$. 
In the MSW regions (both HIGH and LOW), observables depend on 
$\epsilon$ linearly. Obviously the dependence is non-symmetric 
with respect to $\epsilon \leftrightarrow - \epsilon$. 
In the region of strong adiabatic conversion (bottom of the pit)
we get for the survival probaility, the Day-Night asymmetry and the
distortion parameter: 
\begin{equation}
P_{D} \propto (1 -  \epsilon), ~~~~~~
A_{\rm N-D},~ s_{\nu} \propto (1 + \epsilon). 
\label{avprob}
\end{equation}
At the edges of the MSW region (edges of the suppression pit) we find
\begin{equation}
s_{\nu} \propto \epsilon.
\label{slopeed}
\end{equation}
The quasi-vacuum oscillation regions, QVO$_{\rm L}$ and QVO$_{\rm S}$, 
are transition regions between the MSW and vacuum oscillation regions, 
where the linear dependence of observables transform into 
quadratic dependence. 

Thus, for  $\epsilon \ll 1$ the sensitivity  of experiments to 
deviation from maximal mixing is much higher in the MSW regions. 
It will be difficult to measure $\epsilon$ near maximal mixing
if $\Delta m^2$ turns out to be in the VO or QVO regions.

\section{Present Status of Maximal Mixing}
\label{sec:status}

In this section we describe the present status of maximal as well as 
near--maximal mixing. We use for this purpose the latest available 
results on solar neutrinos from Homestake \cite{chlorine}, 
SAGE \cite{sage}, GALLEX+GNO \cite{gallex,gno}, and the 1117 days of 
data sample of Super--Kamiokande \cite{superk,suzuki}\ experiments. 

We calculate the acceptability of maximal and near--maximal mixing
as a function of $\Delta m^2$ in the whole allowed range, that is, below the 
upper bound $\Delta m^2\leq 10^{-3}$  eV$^2$ from the reactor experiments
\cite{CHOOZ,palo}. The goal of this study is to find excluded regions 
of the oscillation parameters $\Delta m^2-\epsilon$, as well as the regions 
of these parameters which are allowed and most plausible. 
We quantify our statements in terms of confidence level 
at which a given region is accepted  (probability of realization) or
excluded. 
We study the dependence of our conclusions on uncertainties in the 
solar neutrino fluxes (the SSM uncertainties) as well as on the procedure 
employed in the analysis.  Some model- and the procedure-independent 
statements are formulated. 
The extent to which the results of this Section hold in a three 
generation framework with $|U_{e3}|\neq0$ is discussed in 
Sec.~\ref{sec:threenus}.

\subsection{Global fit: allowed and forbidden regions}
\label{subsec:globalf}

\begin{figure}[!t]
\centerline{\psfig{figure=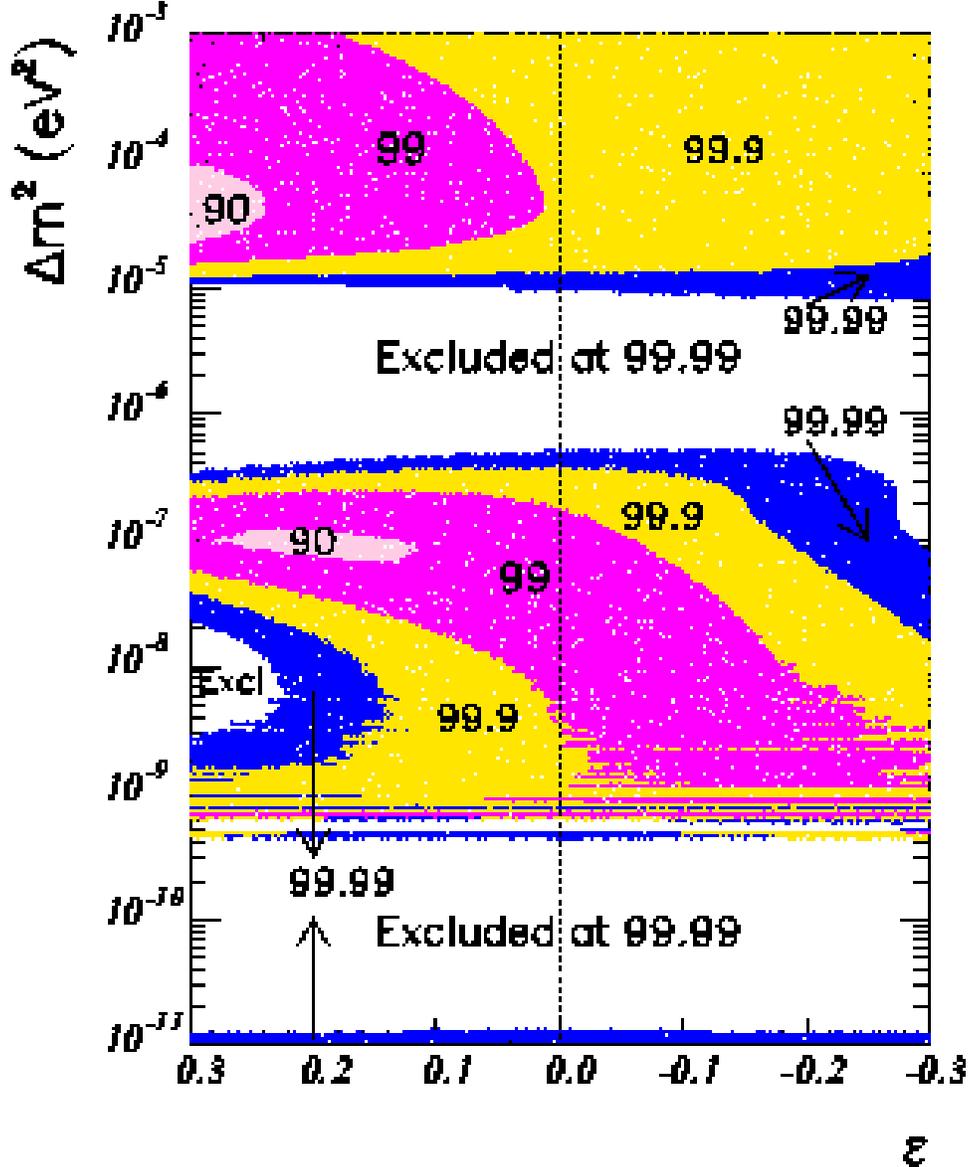,width=5.0in,angle=0}}
\tightenlines
\caption[]{\small 
 Contours of constant confidence level (iso-contours) in the
 $\Delta m^2-\epsilon$ plane. The shaded regions are accepted
 at the given or lower confidence level by the global fit of 
 all available solar neutrino data. The white area is excluded
 at 99.99\% confidence level.
\label{glob}}
\end{figure}

\begin{figure}[!t]
\centerline{\psfig{figure=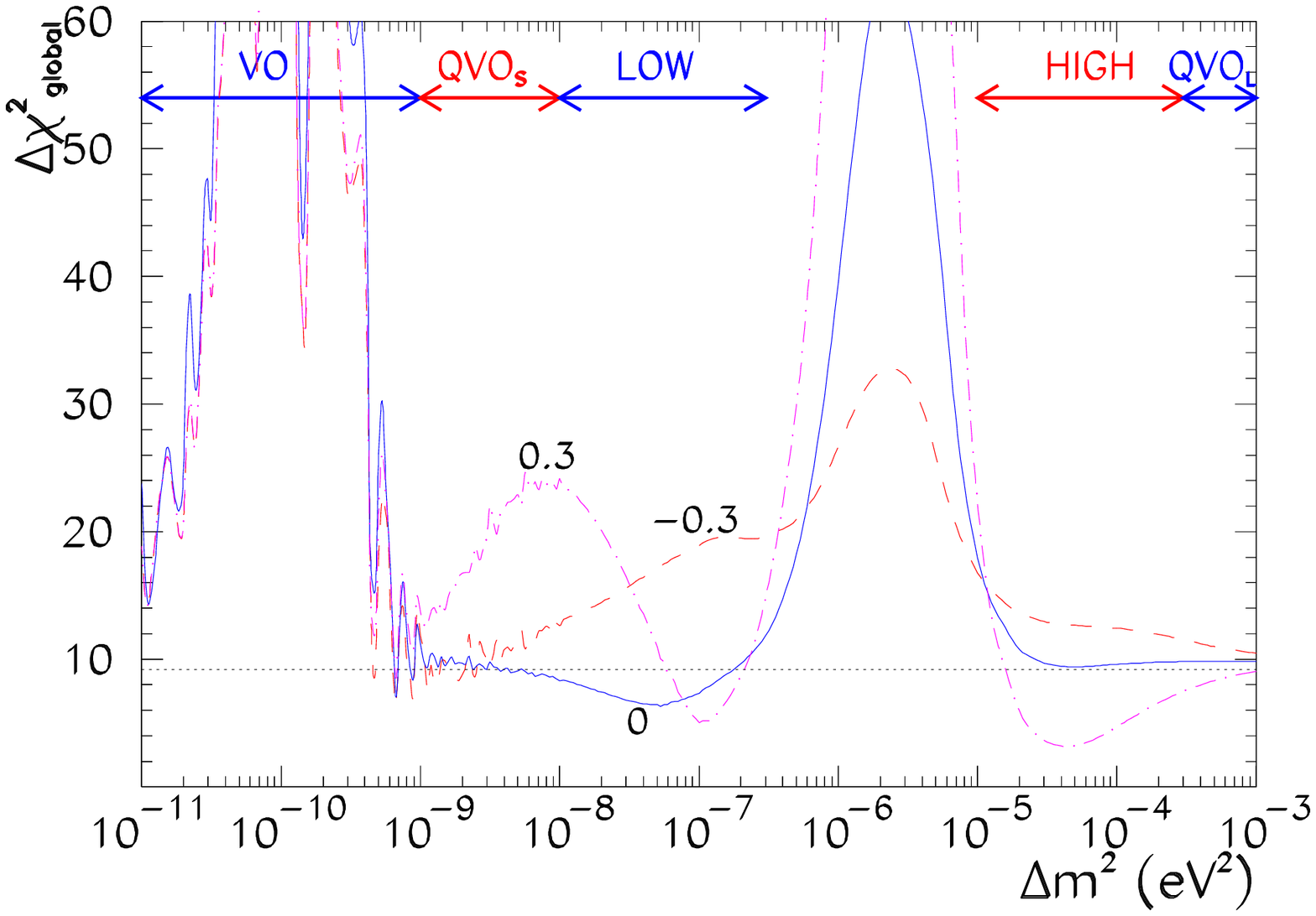,width=5.0in,angle=0}}
\tightenlines
\caption[]{\small The dependence of $\Delta\chi^2$ from the global 
analysis (with the boron neutrino flux of the SSM) on $\Delta m^2$ for 
$\epsilon=-0.3$ (dashed line), $0$ (solid line) and  $+0.3$ (dash-dotted 
line). Dotted horizontal line marks 99\% CL. Below this line the 
corresponding oscillation parameters are accepted at a confidence level 
lower than 99\%. 
\label{chi2_global}}
\end{figure}

\begin{figure}[!t]
\centerline{\psfig{figure=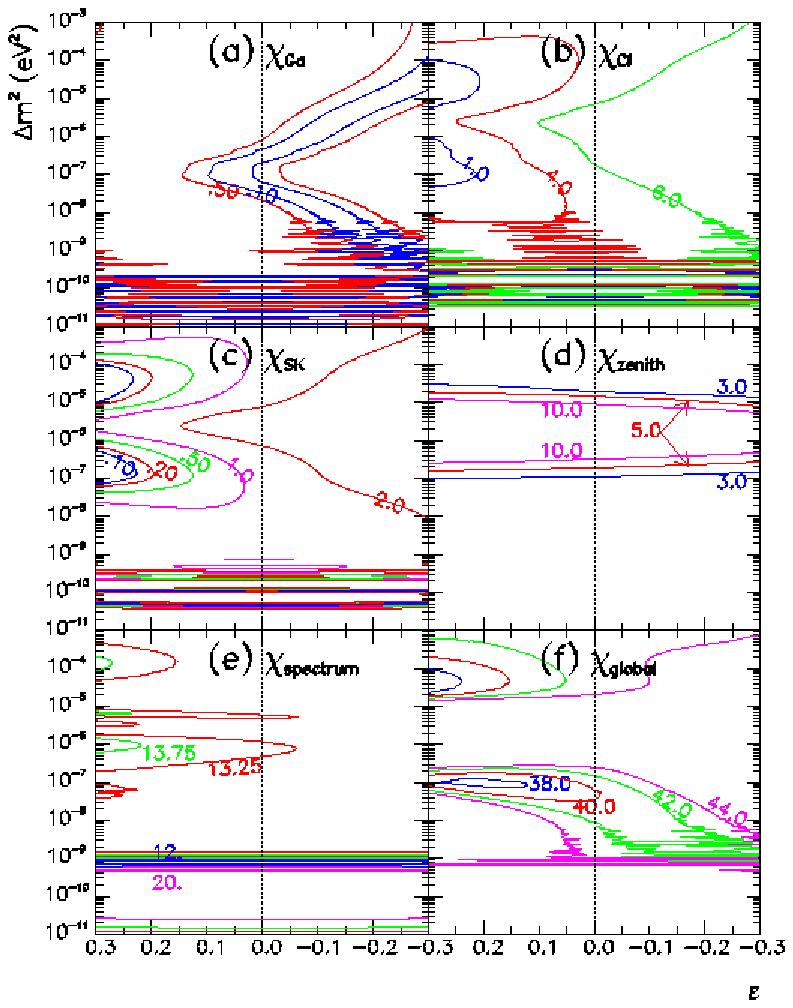,width=5.0in,angle=0}}
\tightenlines
\caption[]{\small Contours of constant $\chi^2$ in the $\Delta m^2-\epsilon$ 
plane for individual experimental results. The various panels show the 
contributions of different results to the total $\chi^2$: (a) total rates in 
SAGE and GALLEX, (b) total rate in Homestake, and the Super--Kamiokande 
measurements of (c) the total rate, (d) zenith angle dependence, and 
(e) recoil electron energy spectrum. In (f) we show iso--contours of the 
total $\chi^2$. 
\label{xic}}
\end{figure}

\begin{figure}[!t]
\centerline{\psfig{figure=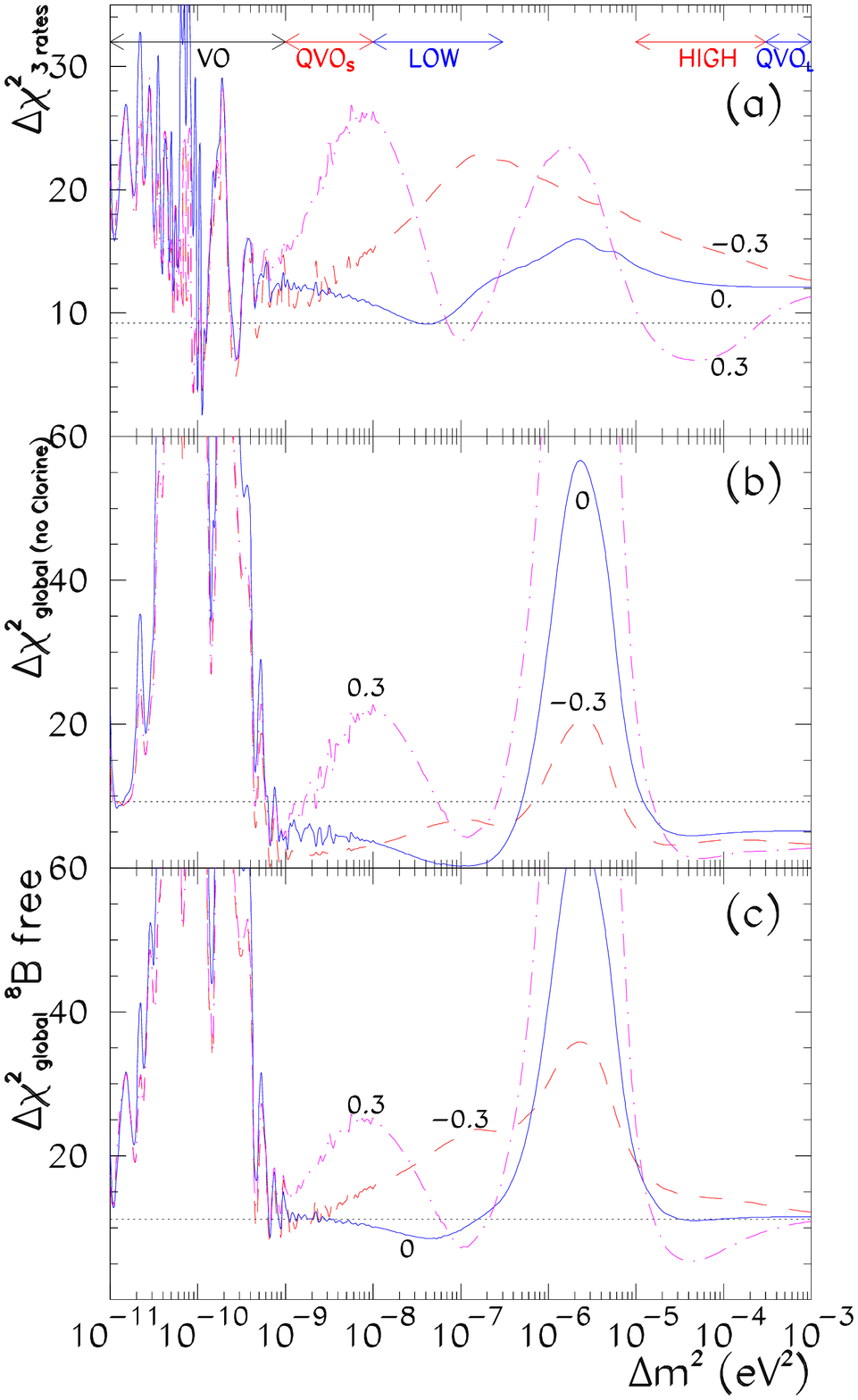,width=5.0in,angle=0}}
\tightenlines
\caption[]{\small The dependence of $\Delta\chi^2$ on $\Delta m^2$ for 
$\epsilon = - 0.3$ (dashed line), $0$ (solid line) and  $0.3$
(dash-dotted line). Dotted horizontal lines mark 99\% CL. Below this line the 
corresponding oscillation parameters are accepted at a confidence level 
lower than 99\%. The three panels correspond to the following fits: 
(a) only total rates; (b) all the data except for the Homestake result; 
(c) all data with the boron neutrino flux treated as a free parameter.  
\label{chi2_part}}
\end{figure}

The results of a global fit to all existing experimental data on solar
neutrinos are shown in Figs.~\ref{glob}$-$\ref{chi2_part}. 
The analysis includes rates in Chlorine \cite{chlorine}, 
Gallium \cite{sage,gallex,gno}, and Super--Kamiokande \cite{superk,suzuki}\
experiments, as well as the zenith angle dependence \cite{superk,suzuki} and 
the shape of the recoil electron spectrum \cite{superk,suzuki} 
in Super--Kamiokande. 

In Fig.~\ref{glob} we plot the contours of constant confidence level 
(iso--contours) in the $\Delta m^2-\epsilon$ plane. Points inside a given 
contour are accepted at a lower confidence level than on the contour itself. 
In the ``global'' analysis we combine the information on the Day--Night 
variation of the event rates and the recoil energy spectrum at 
Super--Kamiokande by using their independently measured spectra during the 
day and during the night. With this the total number of independent 
experimental inputs in the global analysis is 38 which includes 3 rates, and 
35 data points for the Super--Kamiokande day and night recoil energy spectra
($2\times 18$ bins minus 1 overall normalization). We do not include in the
analysis the new lower energy bin as its systematic uncertainty is still under
study by the Super--Kamiokande Collaboration \cite{suzuki}.  We use the solar 
neutrino fluxes from the Standard Solar Model (SSM) of 
Ref.~\cite{bp98}\ (BP98). 
The contours have been defined by the shift 
in $\chi^2$, $\Delta \chi^2$,  
with respect to the global minimum in the whole plane of the oscillation
parameters. The minimum lies in the LMA solution region: 
\begin{equation}
\chi^2 = 33.4\ \ \mbox{\rm for}\ 36\ \ 
\mbox{\rm d.o.f}.  
\label{chimin}
\end{equation}
which corresponds to a probability of 59\%. For details on the 
statistical analysis we refer to Ref.~\cite{gonzalez}.

In Fig.~\ref{chi2_global} we show the dependence of 
$\Delta \chi^2$ on  $\Delta m^2$ for three different values 
of $\epsilon$ ($-0.3,0 ,+0.3$). This figure corresponds to three 
$\Delta m^2$-profiles (cuts) of the confidence level from 
Fig.~\ref{glob}.   


What is the impact of individual experimental results on the global fit? 
Panels (a)$-$(f) of Fig.~\ref{xic} show the contours of constant 
$\chi^2_i$ for each individual observable $i$. 
As mentioned above the total $\chi^2$ in panel (f) is obtained by
combining the $\chi^2$ of the individual rates (including the correlation
of their theoretical errors) with the corresponding $\chi^2$ for the
Super--Kamiokande night and day recoil energy spectra.
 
Our calculations allow us to define regions of oscillation parameters
that are excluded and other that are allowed. It follows from 
Fig.~\ref{glob}\  
that there are two main regions of $\Delta m^2$  which are {\it excluded} at 
a very high (more than $99.99\%$) confidence level: 

1. The regeneration region: for maximal mixing we find  
\begin{equation}
\Delta m^2 = (0.6 - 8) \times 10^{-6}~~ \mbox{\rm eV}^2 ~~~(\epsilon = 0). 
\label{regenreg}
\end{equation}
The excluded region increases with $\epsilon$ for positive values of 
$\epsilon$. In particular, for $\epsilon = 0.3$, the excluded range
becomes    
$\Delta m^2 \sim (0.4 - 10) \times10^{-6}\ \mbox{\rm eV}^2$.   
The excluded region also expands  
for negative values of $\epsilon$ at $\epsilon \lsim -0.25$. 

\begin{figure}[!t]
\centerline{\psfig{figure=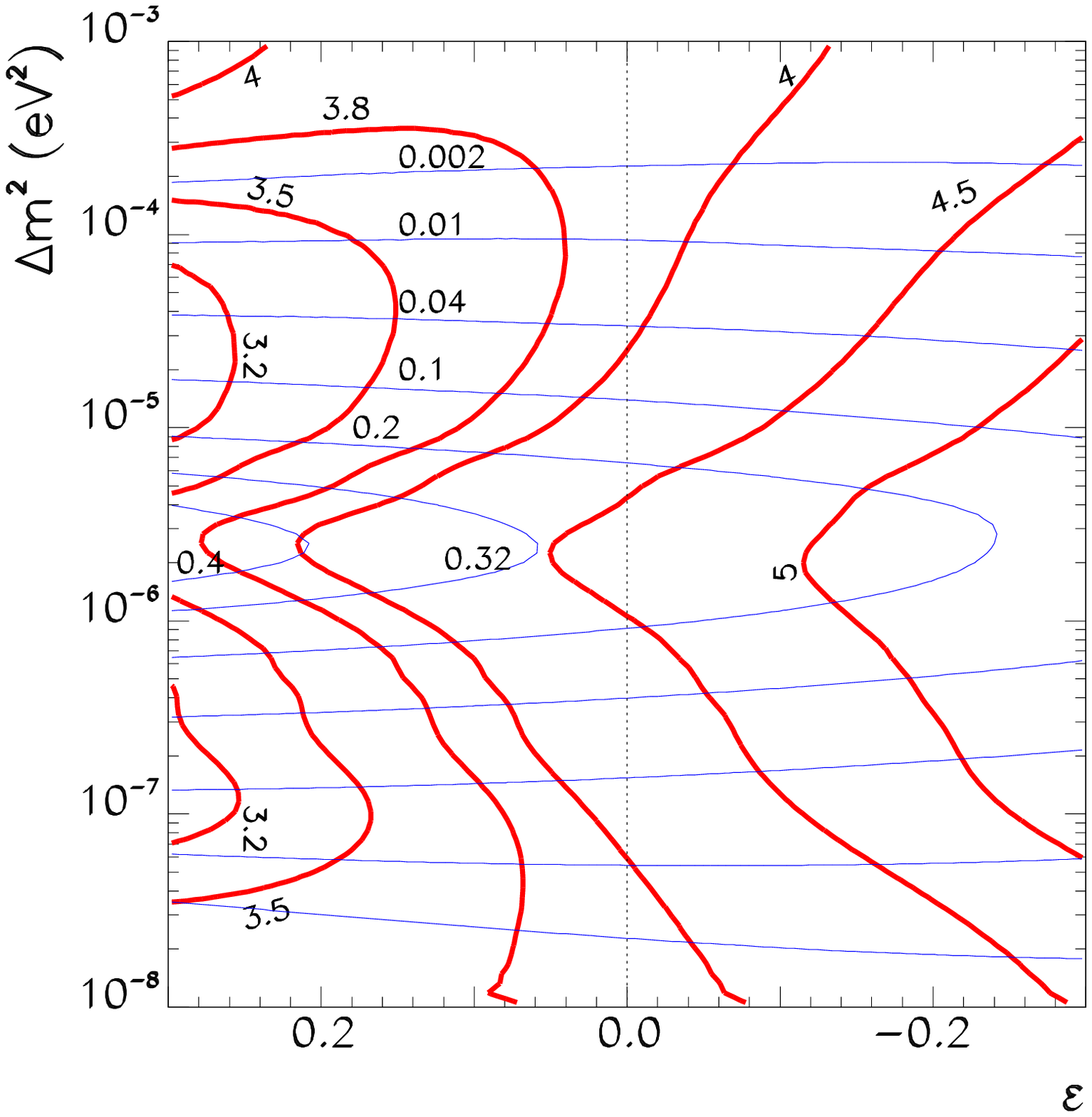,width=5.0in,angle=0}}
\tightenlines  
\caption[]{\small
Contours of constant Day--Night asymmetry at Super--Kamiokande
(thin lines) and of constant Ar--production rate 
(figures at the thick curves in SNU) in the $\Delta m^2 - \epsilon$ plane. 
\label{ardn}}
\end{figure}  

In the regeneration region the solar neutrino observables are strongly
modified by the $\nu_e$-regeneration in Earth. As we have discussed in 
Sec.~\ref{sec:physics}, the regeneration is always positive, thus leading
to an increase in the $\nu_e$-flux. Correspondingly, the counting rates in all 
experiments increase. 
There are two main physical effects in the regeneration region that
are inconsistent with observations and therefore lead to the exclusion:
   
(i) A large Day--Night asymmetry, $A_{\rm N-D} \gsim 0.2$ 
(see Figs.~\ref{xic}(d) and \ref{ardn}), is in contradiction 
to the Super--Kamiokande result and plays a 
dominant role for positive $\epsilon$.

(ii) A large Ar--production rate, $Q_{Ar} > 4$ SNU 
(see  Figs.~\ref{xic}(b) and \ref{ardn}), is in contradiction  
to the Homestake result and leads to the increase of the excluded
region for negative $\epsilon$.  

2. The vacuum oscillation region:
\begin{equation}
\Delta m^2 = (0.14 - 4) \times 10^{-10}~~ \mbox{\rm eV}^2 ~~~(\epsilon = 0).
\label{vac}
\end{equation}
The size of this region depends very weakly  on $\epsilon$ in the interval 
$-0.3\lsim\epsilon\lsim+0.3$. The exclusion  follows from the interplay
between the total rates and the shape of the recoil electron energy 
spectrum. Notice that the rates only (see Fig.~\ref{chi2_part}(a)) 
can be reproduced rather well in some parts of the region in Eq.~(\ref{vac}).  
However, in these regions the distortion of the recoil electron spectrum is 
in contradiction with the Super--Kamiokande results (see Fig.~\ref{xic}(e)). 
More specifically, for $\Delta m^2 \sim 10^{-10}$ eV$^2$, a negative slope 
(or shift of the first moment) in the ``reduced" spectrum is expected. 
The reduced spectrum is defined as the ratio 
\begin{equation}
R(E)\equiv N_{i}(E)/N_{i}(E)^{{\rm SSM}},
\label{defRE}
\end{equation}
where $N_{i}(E)$ ($N_{i}(E)^{\rm SSM}$) is the number of events in a 
given electron energy, $E$, bin $i$ with (without) oscillations. 
In the same way we
define any ``reduced'' observable as the ratio of its value with 
respect to the expected one in the SSM in the absence of oscillations. 

3. There is another region forbidden  at 99.99 \% CL extending from 
$\Delta m^2=(0.3 - 2) \times 10^{-8}~~ 
\mbox{\rm eV}^2$ for $\epsilon = 0.3$
to $\Delta m^2=6 \times 10^{-9}~~\mbox{\rm eV}^2$   
and $\epsilon = 0.22$. As seen in Fig.~\ref{chi2_global} this region
is forbbiden with a lower CL. The exclusion follows mainly from the
effect of the rates as seen in Fig.~\ref{chi2_part}(a) being mainly
driven by the bad fit to the Gallium rate (see Fig.~\ref{rates}(a)).

We distinguish five regions of the oscillation parameters where maximal
mixing is {\it allowed} at a confidence level that is lower than 99.9\%: 

1) Quasi vacuum oscillation region with large  $\Delta m^2$ 
(QVO$_{\rm L}$): 
\begin{equation}
\Delta m^2 \sim\left(3\frac{E}{\rm 10\ MeV}- 8\right)\times 10^{-4}\ 
\mbox{\rm eV}^2 , 
\label{qvol}
\end{equation} 
where $E$ is the average detected energy for a given experiment. 
The upper bound comes from reactor experiments \cite{CHOOZ,palo}. 
Here the flavour conversion is mainly due to averaged vacuum oscillations 
with only small matter corrections inside the Sun and the Earth.

2) MSW region with high $\Delta m^2$ (HIGH): 
\begin{equation}
\Delta m^2 \sim  \left(0.1 - 3 \frac{E}{\rm 10\ MeV}\right) 
\times 10^{-4}\ \mbox{\rm eV}^2 . 
\label{high}
\end{equation}
This region corresponds to the maximal and near--maximal mixing part of the 
LMA solution. It is restricted from below by strong Earth regeneration
effects (large Day--Night asymmetry and large Ar--production rate). 
Maximal mixing is acceptable at confidence level larger than  $99.1 \%$. 
As follows from 
Fig.~\ref{chi2_global}, the dependence of $\Delta \chi^2$ on $\Delta m^2$ 
is rather weak. The global fit becomes substantially better with increase 
of $\epsilon$ (shift to positive values): $\epsilon\gsim0.25$ is accepted 
at 90\% CL. For $\epsilon=0.3$, the 90\% CL allowed region expands to
$\Delta m^2 = (2 - 10)\times 10^{-5}$  eV$^2$. In contrast, the goodness 
of the fit decreases when we shift to  negative values of $\epsilon$. 
 
3) MSW region with low $\Delta m^2$ (LOW): 
\begin{equation}
\Delta m^2 \sim  (0.1- 3) \times  10^{-7}\ \mbox{\rm eV}^2.
\label{low}
\end{equation}
Here maximal mixing is acceptable at $\sim 99\%$ CL in the interval 
\begin{equation}
\Delta m^2 \sim (0.1 - 2) \times 10^{-7}\ \mbox{\rm eV}^2\ 
(\mbox{\rm CL} \leq 99\%).  
\label{lowbest}
\end{equation}
For positive $\epsilon$ the fit improves while for negative $\epsilon$
it worsens. In particular,  $\epsilon =0.2$ is accepted at $\sim
90\%$ CL for $\Delta m^2\sim (0.8-1.5)\times 10^{-7}$ eV$^2$. 
The local minimum occurs at $\Delta m^2= 1.0 \times 10^{-7}\ 
\mbox{\rm eV}^2$ and $\epsilon=0.21$. 
With increase of $\epsilon$ the accepted region of $\Delta m^2$ shifts
to larger values. At $\epsilon= 0.3$ we obtain
$\Delta m^2 = (0.6-2) \times 10^{-7}\ \mbox{\rm eV}^2$.
Conversely negative values of $\epsilon$ are disfavored. 

4) Quasi vacuum oscillation region with small $\Delta m^2$ (QVO$_{{\rm S}}$):
\begin{equation}
\Delta m^2 \sim  (0.1- 1) \times 10^{-8}\ \mbox{\rm eV}^2.
\label{qvos}
\end{equation}
In this region the flavour conversion is due to (mainly non-averaged) vacuum 
oscillations with small matter effects. 
Maximal mixing is acceptable at $\sim 99\%$ CL in the interval 
\begin{equation}
\Delta m^2 \sim (0.5 - 1) \times 10^{-8}\ \mbox{\rm eV}^2\ 
(\mbox{\rm CL} \sim 99\%).
\label{qvolbest}
\end{equation}
The fit becomes worse with increase of $|\epsilon|$, 
but while for $\epsilon>0$ the QVO$_{{\rm S}}$ region 
is essentially excluded, 
for $\epsilon < 0$ we still have a reasonably good fit. 

5) Vacuum oscillation region with relatively large $\Delta m^2$ 
(VAC$_{{\rm L}}$):   
\begin{equation}
\Delta m^2 \sim(0.4-1)\times10^{- 9}~~ \mbox{\rm eV}^2.   
\label{vacl}
\end{equation}
Maximal mixing is accepted at a confidence level better than 99\%  
only in a very small interval centered at:  
\begin{equation}
\Delta m^2 \sim 6.6 \times 10^{-10}\ \mbox{\rm eV}^2~~~(\mbox{\rm CL} \leq 99\%).\label{vacbest}
\end{equation} 
In this interval, the goodness of the fit depends on $\epsilon$ very weakly.

Summarizing, maximal (or near--maximal) mixing is allowed 
at $99\%$ or slightly lower CL in several small intervals of 
$\Delta m^2$ in the QVO$_{\rm L}$, HIGH, LOW, QVO$_{\rm S}$  
and VAC$_{\rm L}$ solution domains. 
The values $\epsilon = 0.05,~ 0.1, ~ 0.2$   are allowed at 99, 95 and 
90\% CL, respectively. At $4 \sigma$,  practically the whole HIGH, 
LOW, QVO and VAC$_{{\rm L}}$ ranges are allowed.  

Let us point out the role of individual experimental results in 
constraining maximal mixing (see Fig.~\ref{xic}). The rates in the Gallium and
Super--Kamiokande 
experiments can be well accounted for at maximal (or near--maximal) mixing, 
although the Super--Kamiokande measurement slightly disfavors a negative 
$\epsilon$. The zenith angle distribution measured by Super--Kamiokande 
gives some preference to the HIGH region and excludes the strong regeneration 
region. In contrast, the Super--Kamiokande result on the recoil electron 
energy spectrum gives some preference to the VAC$_{\rm L}$, HIGH and LOW  
regions and excludes the range 
$\Delta m^2 = (0.3 - 4) \times 10^{-10}\ \mbox{\rm eV}^2$. 
Both the zenith angle distribution and the shape of 
spectrum have weak dependence on $\epsilon$. In contrast, total rates
are sensitive to $\epsilon$, especially in the HIGH and LOW regions.

\subsection{Dependence of the results on features of the analysis}
\label{subsec:dependence}
Let us study the dependence of the allowed and excluded regions in
the $\Delta m^2-\epsilon$ plane on features of our analysis.  

{\it Total rates versus spectrum and zenith angle distribution:} 
The total rates give the most stable, reliable, and statistically
significant information. 
We have carried out a fit to the three rates only.
In Fig.~\ref{chi2_part}(a) we show the dependence of the shift of  
$\chi^2$ for this analysis with respect to the absolute minimum in the whole 
plane of oscillation parameters. The absolute minimum for the analysis
of the three rates, $\chi^2=0.76$ (for one
d.o.f) is achieved in the SMA region. Comparing  Fig.~\ref{chi2_part}(a)
with Fig.~\ref{chi2_global}, we learn that if we exclude the
information from the recoil electron energy spectrum and the Day--Night 
variation of the event rates from our analysis, then the 
allowed and forbidden regions are substantially modified. 
In particular, we note in Fig.~\ref{chi2_part}(a) the following three 
features: 

(i) The goodness of the fit at maximal mixing from the analysis of
the three rates only is worse in whole MSW 
region. At 99\% CL only  small interval in the LOW region is allowed. 
The fit improves however with increase of $\epsilon$. 
This means that  it is the data on the spectrum and the zenith angle 
distribution which favor maximal mixing. 

(ii) Allowed regions appear in the VAC solution range. 
We learn that the data on 
the spectrum and the zenith angle distribution exclude (otherwise) allowed 
VAC regions.
 
(iii) The regeneration region is still strongly disfavored by the high 
Ar--production rate. \\

{\it The Homestake result:} 
Consider the impact of the Ar--production rate on our results. 
In Fig.~\ref{xic}(b) we show the fit to {\it only} this rate. From 
the figure we see that the Homestake result strongly disfavors maximal 
mixing for all 
$\Delta m^2$ above $10^{-10}$ eV$^2$, that is, in all the globally 
allowed  regions. In Fig.~\ref{chi2_part}(b) we show the result of 
a global fit to the data {\it without} the Homestake result. Clearly, 
the acceptability of maximal mixing improves for all $\Delta m^2\gsim
10^{-10}$ eV$^2$ with the best fit points being in the LOW region, 
$\Delta m^2 = 10^{-7}\ \mbox{\rm eV}^2$. For 
$\epsilon = 0.3$  the best fit is  in VAC$_{\rm L}$ region.  
We would like to emphasize the following points about a fit without
the Homestake result:

(i) In the HIGH region, maximal mixing is accepted already at 1.7 $\sigma$ 
with very little dependence on $\epsilon$.

(ii) In the whole LOW region, maximal mixing gives a very good fit.

(iii) In the whole QVO$_{{\rm L}}$ region, maximal mixing gives 
a very good fit, but positive values of $\epsilon$ are still disfavored.   
 
(iv) Regions of strong regeneration and VAC (small $\Delta m^2$) solutions 
are  excluded by the Super--Kamiokande data on the spectrum  
and the Day--Night asymmetry. 

{\it Solar neutrino flux uncertainties:} 
Of all the relevant solar neutrino fluxes, the boron neutrino flux suffers 
from the largest uncertainty, leading to systematic errors in the 
predicted detection rate that cannot be estimated reliably at present. 
One way to avoid this problem is to determine the boron neutrino flux 
experimentally, using the total rate measured in the Super--Kamiokande 
experiment. 
Similar results are obtained if the boron neutrino flux is treated as 
a free parameter in the analysis (in this case the Super--Kamiokande rate,  
being the most precise and sensitive to the boron neutrino flux,  
will fix this flux). In Fig.~\ref{chi2_part}(c) we show the results of 
a global fit with the boron neutrino flux treated as a free parameter. 
We plot the dependence of $\Delta \chi^2$, the $\chi^2$ shift with 
respect to the absolute minimum, on $\Delta m^2$ for three different 
values of $\epsilon$. The shape of the curves is very similar to that in 
Fig.~\ref{chi2_global}. Also the allowed and excluded regions at a given 
CL practically coincide with those in  Fig.~\ref{chi2_global}. 
(One should take into account that now we have one additional free
parameter and therefore a 99\% CL corresponds to $\Delta\chi^2=11.36$). 
The reason of this similarity lies on the fact that both the spectrum
and the Day--Night variation are flux independent.
Furthermore in both cases the boron neutrino flux is fixed by the
Super--Kamiokande result. 

A comparison of Fig.~\ref{chi2_global} and Fig.~\ref{chi2_part}(c)
shows that the results of our analysis are stable with respect to the way 
that the uncertainty in the boron neutrino flux is treated. 
 
\section{Maximal Mixing and Predictions for Individual Experiments} 
\label{sec:predic}
In this section we consider the predictions of (near--) maximal mixing 
for various observables. This will clarify the sensitivity of individual 
experiments to the neutrino oscillation parameters in the relevant range. 
The extent to which the results of this Section hold in a three 
generation framework with $|U_{e3}|\neq0$ is discussed in 
Sec.~\ref{sec:threenus}.

\subsection{Total rates}
\label{subsec:rates}
\begin{figure}[!t]
\centerline{\psfig{figure=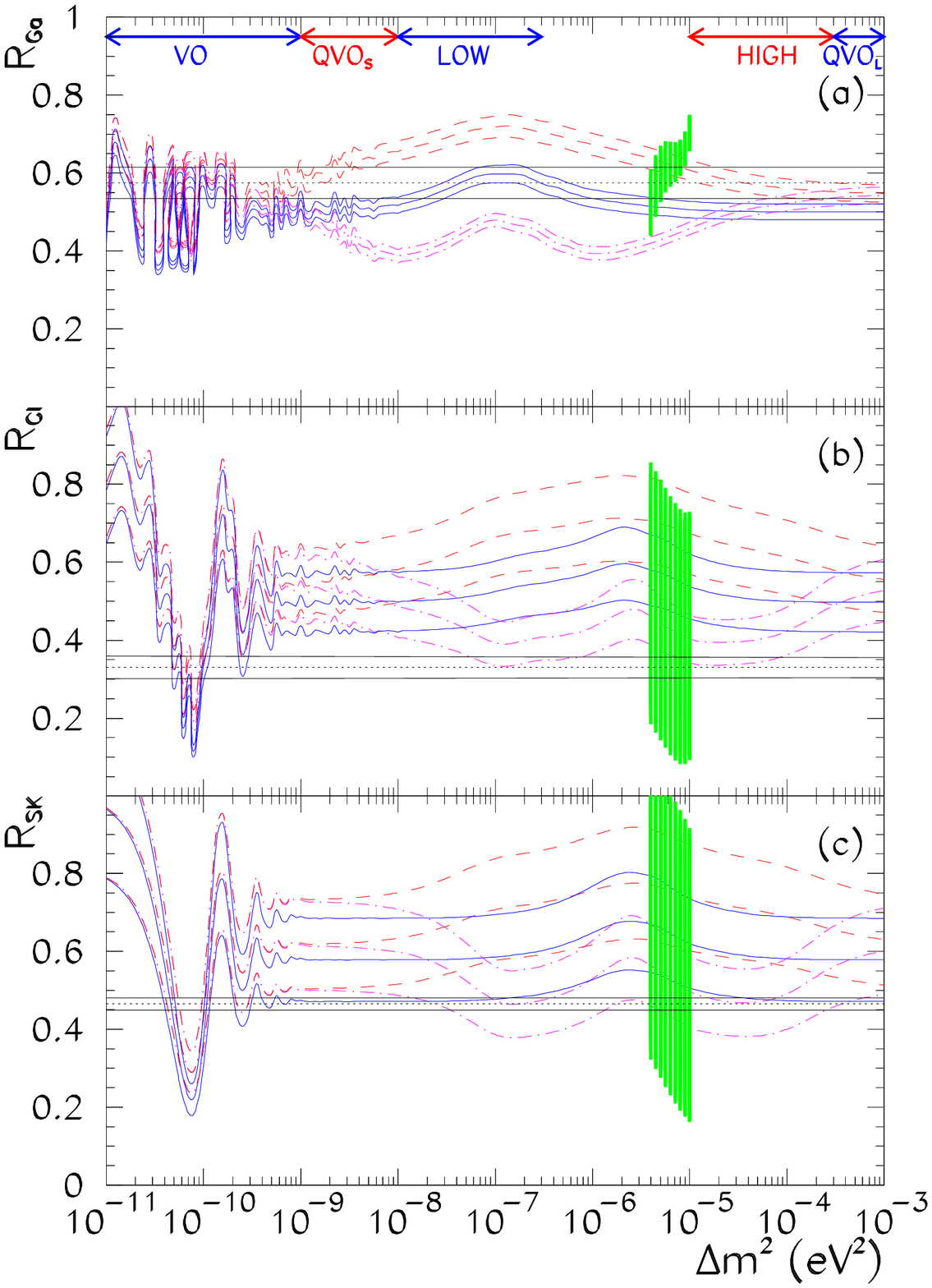,width=5.0in,angle=0}}
\tightenlines  
\caption[]{\small (a) The Germanium production rate, (b) the Argon production 
rate and (c) the rate of events at Super--Kamiokande, as functions of 
$\Delta m^2$ for 3 values of $\epsilon$: $+0.3$ (dash-dotted lines), 
$0$ (solid lines), $-0.3$ (dashed lines). The rates are 
normalized  to the BP98 no oscillation expectation. Also shown are
the experimental constraints and the SMA prediction. See the text for
more details.
\label{rates}}
\end{figure}  

In  Fig.~\ref{rates} we plot the values of the expected event rates: 
Germanium production rate, Argon production rate, and the rate of events at 
Super--Kamiokande as functions of $\Delta m^2$ for 3 values of
$\epsilon$: $+0.3, 0,-0.3$. The rates are normalized  to the 
no oscillation expectation, $R_i\equiv Q_i/Q_i^{{\rm SSM}}$. 
For each value of $\epsilon$ we plot three curves: the central curves 
give the expected rates using central values of the BP98 fluxes and the 
upper and lower lines represent the theoretical 
uncertainty (without the error for the interaction cross sections)
from varying
the nine parameters in the SSM within $\pm 1\sigma$. The horizontal 
lines give  the experimental values within $\pm 1\sigma$  experimental 
errors. The vertical lines in the range $\Delta m^2=(3- 10)\times
10^{-6}$  eV$^2$ give the expectation from  the SMA 99\% CL region (again,
including the theoretical uncertainties in each point). 

Let us first discuss the dependence of the rates on $\Delta m^2$. 
The rates are proportional to the survival probability. Therefore, 
the main features of Fig.~\ref{rates} reflect  the dependence 
of the survival probability on $\Delta m^2/E$ (see Fig.~\ref{fig:sun}
and Eqs.~(\ref{pdepsL}), (\ref{pdepsI}) and (\ref{avprob2})). 
For maximal mixing the survival probability in the Sun 
as a function of $\Delta m^2/ E$ is constant:  $P_D=1/2$
(see Eq.~(\ref{Psubd})). The probability 
$P_{ee}$ is  enhanced by the Earth regeneration effect for $\Delta m^2/E$
in the range $(10^{-15} - 10^{- 11})\ \mbox{\rm eV}$ 
(see Eq.~(\ref{barP})). For $\epsilon\neq 0$, 
the  effects of the adiabatic edge situated at 
$\Delta m^2/E=(10^{-12}-10^{- 10})$
eV (Eq.~(\ref{adiabprob1})), and of the non-adiabatic edge situated 
at $\Delta m^2/E=(10^{-16}-
3\times10^{- 15})$ eV (Eq.~(\ref{adiabprob2})), become important. 
For $\Delta m^2/E\lsim10^{-16}$ eV,
an oscillatory behaviour appears due to non-averaged vacuum oscillations 
between the Sun and the Earth (Eq.~(\ref{avprob2})). With certain 
modifications all these features 
can be seen in Fig.~\ref{rates}. The simplest dependence is for the 
Super--Kamiokande rate since only one (boron) neutrino flux contributes 
(Fig.~\ref{rates}(c)). 
The Ar--production rate has an additional fine structure due to contributions
from additional fluxes, and in particular the beryllium neutrino flux. 
As can be seen from Fig.~\ref{rates}(b), an additional enhancement appears
in the regeneration region at  $\Delta m^2=3\times10^{-7}\ 
\mbox{\rm eV}^2$ and  
the probability as a function of $\Delta m^2$  becomes asymmetric in the 
regeneration region. In the VAC and QVO$_S$ regions the boron neutrino
``wave" 
is modulated by the beryllium wave with a smaller amplitude. For the 
Ge--production rate, all the features of the curves are shifted 
(with respect to the Super--Kamiokande curves) to smaller values of 
$\Delta m^2$ by a factor $\sim 30$.  
This feature is due the fact that the main contribution to $Q_{Ge}$ comes 
from the $pp$-neutrino flux with an average detected energy of 0.3 MeV, 
about 30 times smaller than the  average energy of the boron neutrino flux.  

Let us consider now the $\epsilon$-dependence of the rates. We distinguish
here between three different regions of $\Delta m^2$: 

1)  The QVO$_{{\rm L}}$  region with $\Delta m^2\gsim 3\times10^{-4}\ 
\mbox{\rm eV}^2$. 
Here we have essentially averaged vacuum oscillations, so that the
survival probability is given by Eq.~(\ref{pdepsL}).
 
2) The QVO$_{\rm S}$ and VAC regions with $\Delta m^2\lsim10^{-8}\ 
\mbox{\rm eV}^2$. Here  
we have essentially non-averaged or partially averaged vacuum oscillations
which take the form of Eq.~(\ref{avprob2}).
In principle matter effects strongly suppress the oscillations inside 
the Sun and the Earth. 
However, the modification of the observables is small, since the size of 
the Sun  (and the Earth) is much smaller than the oscillation length in 
vacuum. 

3) The matter conversion (MSW) region is between these two QVO regions.   
The oscillation effect is strongly suppressed. As explained in 
Sec.~\ref{sec:physics}, Earth matter effect is important but
insensitive to $\epsilon$. The expression of $P$ in this region is
given in Eqs.~(\ref{adiabprob1}) and (\ref{adiabprob2}).

From Eqs.~(\ref{pdepsL}), (\ref{pdepsI}), and (\ref{avprob2}) 
we can find straightforwardly the dependence of the observables on 
deviations from maximal mixing. For the VO, QVO$_{{\rm S}}$ and 
QVO$_{{\rm L}}$ regions, the following points are in order:
\begin{enumerate}
\item  The observables depend very weakly on $\epsilon$. Corrections 
to maximal mixing are of ${\cal O}(\epsilon^2)$. 
\item The dependence is symmetric under the exchange 
$\epsilon\leftrightarrow-\epsilon$. Minimum of the survival probability, 
and consequently, minima of rates, are at $\epsilon = 0$, that is, 
at maximal mixing. 
\end{enumerate} 

For the MSW regions, the following points are in order:
\begin{enumerate}
\item The survival probability and the rates depend linearly on 
$\epsilon$. Corrections to the maximal mixing case are consequently larger. 
\item The survival probability and the rates decrease with increase 
of $\epsilon$. 
\end{enumerate} 

There are two transition regions between VO and pure matter 
conversion. In these regions, the symmetric dependence of the observables
transforms into a linear dependence and the sensitivity to deviation from 
maximal mixing increases. The ambiguity $\epsilon\leftrightarrow-\epsilon$ 
disappears. 

Thus, in the region of pure matter conversion (inside the Sun) the 
sensitivity of measurements of rates to a deviation from maximal 
mixing is maximal. It is in this region that the possibility of maximal
mixing can be tested with the highest accuracy. The corresponding 
$\Delta m^2$ range depends on the neutrino energy. For the highest 
energies of the solar neutrinos ($E \sim 10$ MeV) we get the range of 
maximal sensitivity:  $\Delta m^2 =  10^{- 7} -  10^{- 4}~\mbox{\rm eV}^2$, 
while for the lowest energies ($E \sim 0.3$ MeV) the range is $\Delta m^2 = 
3\times 10^{- 9} - 3 \times 10^{- 6}~\mbox{\rm eV}^2$. 
Studying effects by the experiments with different energy thresholds
we can get high sensitivity to deviations from maximal mixing in whole
range of $\Delta m^2$ excluding VO. 

In the next two subsections we consider the dependence of specific rates on
the oscillation parameters  and evaluate the sensitivity of their present
measurements to deviations from maximal mixing.

\subsection{Ar--production rate.} 
\label{subsec:argon}
Let us now study the implications of the results presented in 
Fig.~\ref{rates}(b) for the Ar--production rate and in 
Fig.~\ref{rates}(c) for the Super--Kamiokande rate in the 
various $\Delta m^2$ regions. As can be seen in Fig.~\ref{rates}(b), 
the Ar--production rate for maximal mixing in all favorable 
regions (HIGH, LOW, VAC$_{{\rm L}}$) lies in the 
range  $R_{Cl}=0.50 \pm 0.08$  ($Q_{Ar}= 3.9 \pm 0.6$ SNU), which is 
$2\sigma$ above the Homestake result. The predictions for 
$\epsilon=+0.3$ ($- 0.3$) are below (above) the values for maximal mixing 
by just about $1\sigma$. As mentioned above, 
the highest sensitivity of $Q_{Ar}$ to $\epsilon$ is achieved already 
in the HIGH region, where $Q_{Ar}$ depends on $\epsilon$ linearly. 
For the Super--Kamiokande rate, we get that for maximal 
mixing in all favorable regions $R_{SK} = 0.58 \pm 0.12$  
which is $1\sigma$ above the Super--Kamiokande result. Moreover 
comparing Figs.~\ref{rates}(b) and ~\ref{rates}(c) we see that 
the Ar--production rate and the Super--Kamiokande rate are 
strongly correlated as they are both dominated by the contribution from 
boron flux neutrinos. 

\begin{figure}[!t]
\centerline{\psfig{figure=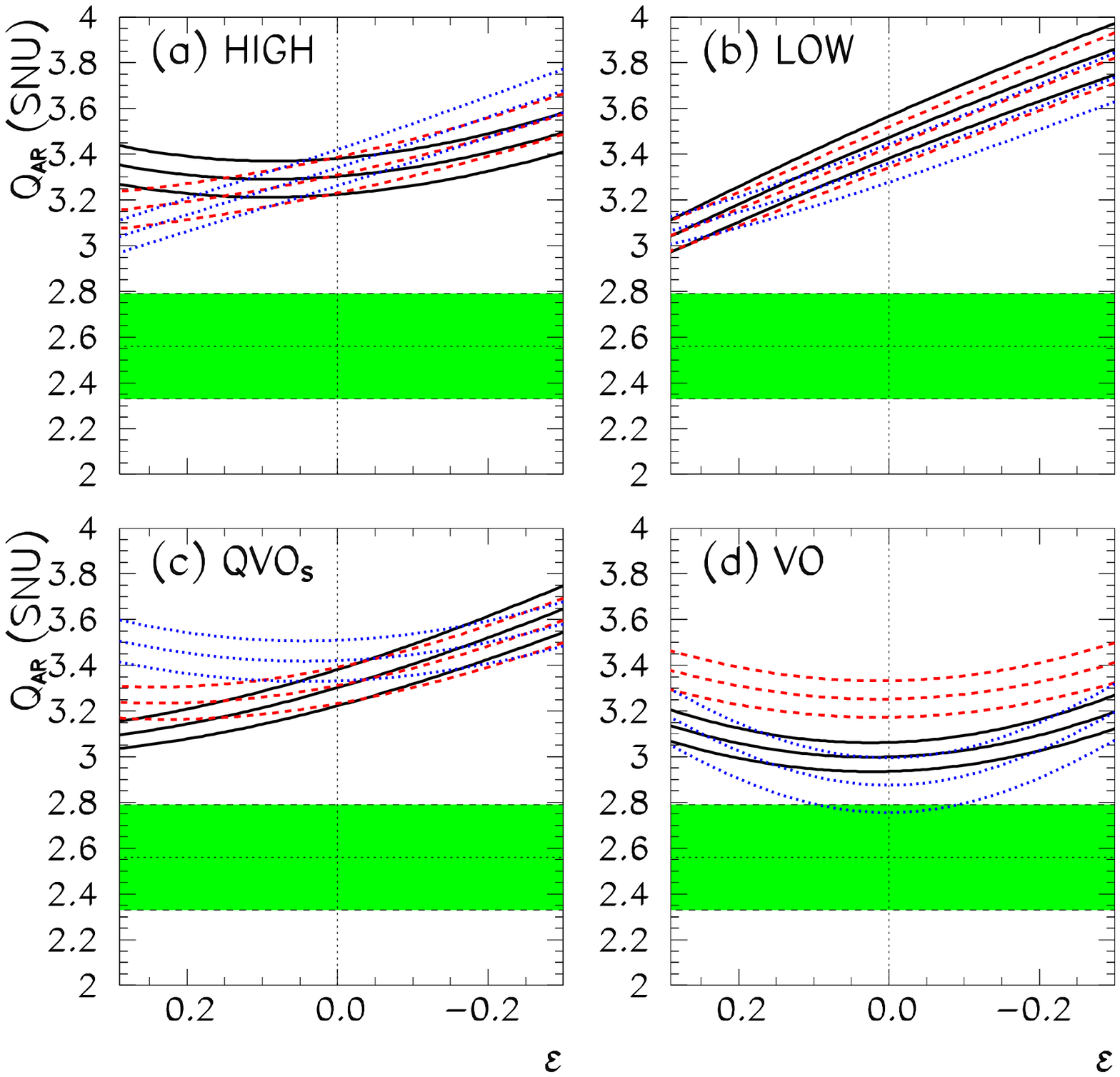,width=5.0in,angle=0}}
\tightenlines  
\caption[]{\small
The dependence of the Ar--production rate on $\epsilon$  for various 
values of $\Delta m^2$ within the (a) HIGH, 
(b) LOW, (c) QVO$_{{\rm S}}$ and (d) VO regions. 
Full, dashed and dotted lines correspond to: 
(a) $\Delta m^2=4,\ 0.8,$ and $0.2 \times 10^{-5}$ eV$^2$,    
(b) $\Delta m^2=1,\ 0.7,$ and $0.3 \times 10^{-7}$ eV$^2$,    
(c) $\Delta m^2=10,\ 5,$ and $1 \times 10^{-9}$ eV$^2$,    
(d) $\Delta m^2=5,\ 0.3,$ and $1 \times 10^{-10}$ eV$^2$, respectively.    
For each mass the three curves correspond to the predicted central 
value plus and minus 1$\sigma$ theoretical uncertainties.
The boron neutrino flux is normalized to the measured 
rate at Super--Kamiokande. The horizontal band corresponds 
to the Homestake experimental result and its 1$\sigma$ error. 
\label{q_ar}}
\end{figure}  

Next, in order to reduce the SSM uncertainty we normalize the boron 
neutrino flux to the Super--Kamiokande rate. More precisely, for each pair 
of the of oscillation parameters ($\Delta m^2,\epsilon$) we find the boron 
neutrino flux which reproduces the Super--Kamiokande event rate. All other 
fluxes and their uncertainties are taken according to the BP98. Using 
this procedure we calculate $Q_{Ar}$ (and in the next subsection also 
$Q_{Ge}$). The results of this calculation are shown in Fig.~\ref{q_ar}. 

From Fig.~\ref{q_ar} we see that after boron flux   normalization
procedure 
the dependence of $Q_{Ar}$ on $\epsilon$ is relatively weak since the 
ratio between the boron neutrino flux contribution to $Q_{Ar}$ and the 
contribution from charged current interactions to the Super--Kamiokande 
rate is independent of the survival probability and therefore 
of $\epsilon$. The $\epsilon$ dependence of $Q_{Ar}$ comes directly from 
the suppression of the beryllium and other (CNO and pep) neutrino fluxes, and  
indirectly from the contribution 
of neutral current interactions to the Super--Kamiokande rate. 
Expressing the boron neutrino flux $f_B$ via the SK rate:  
\begin{equation}
f_B \sim \frac{ R_{SK}}{P(1 - r) + r}, 
\label{fb}
\end{equation}
we obtain:  
\begin{equation} 
Q_{Ar} \sim \frac{Q_{Ar}^{B} R_{SK}}{(1 - r + r/P)} + Q_{Ar}^{\not B} P'.  
\label{arg}
\end{equation}
Here $r$ is the ratio of the $\nu_{\mu}-e$ and  $\nu_e-e$ cross-sections, 
$Q_{Ar}^{B}$ and $Q_{Ar}^{\not B}$ are the SSM contributions to the 
Ar--production rate from the boron neutrino flux and from all other 
low energy fluxes, respectively, and 
$P$ and $P^\prime$ are the effective survival probabilities for
the boron neutrino and low energy neutrino fluxes, respectively.
The bands in Fig.~\ref{q_ar}\ 
reflect $1\sigma$ errors due to the uncertainties in all, but the boron, neutrino fluxes. 

As expected, in the QVO regions, QVO$_{{\rm L}}$ and 
QVO$_{{\rm S}}$, 
the $\epsilon$-dependence is symmetric around $\epsilon=0$, 
$Q_{Ar} = Q_{Ar}(\epsilon^2)$, while in the pure matter conversion regions, 
HIGH and LOW,  $Q_{Ar}$ depends on $\epsilon$ linearly. In the linear regime,
the change in the Ar--production rate is $\Delta Q_{Ar} \sim (0.7 - 0.8)$ SNU 
for  $- 0.3 \lsim\epsilon\lsim+ 0.3$. This is about $3 \sigma$ for the 
present experimental error. In the quadratic regime the change 
is substantially 
smaller: $\Delta Q_{Ar} \sim (0.1 - 0.2)$ SNU. Clearly, the present 
sensitivity is not enough to draw definite conclusions. 
Moreover, even after normalization of the boron neutrino flux to 
the Super--Kamiokande rate, the predicted rate is higher than the 
Homestake result 
for all globally allowed values of $\Delta m^2$. The only statement that 
one can make is that the Homestake result favors a significant deviation 
from maximal mixing in the MSW region. Therefore checks of the Homestake 
result and improved accuracy in measurements of $Q_{Ar}$ by a factor of two 
or higher would have important implications for maximal and near--maximal 
mixing.

\subsection{Ge--production rate} 
\label{subsec:germanio}
\begin{figure}[!t]
\centerline{\psfig{figure=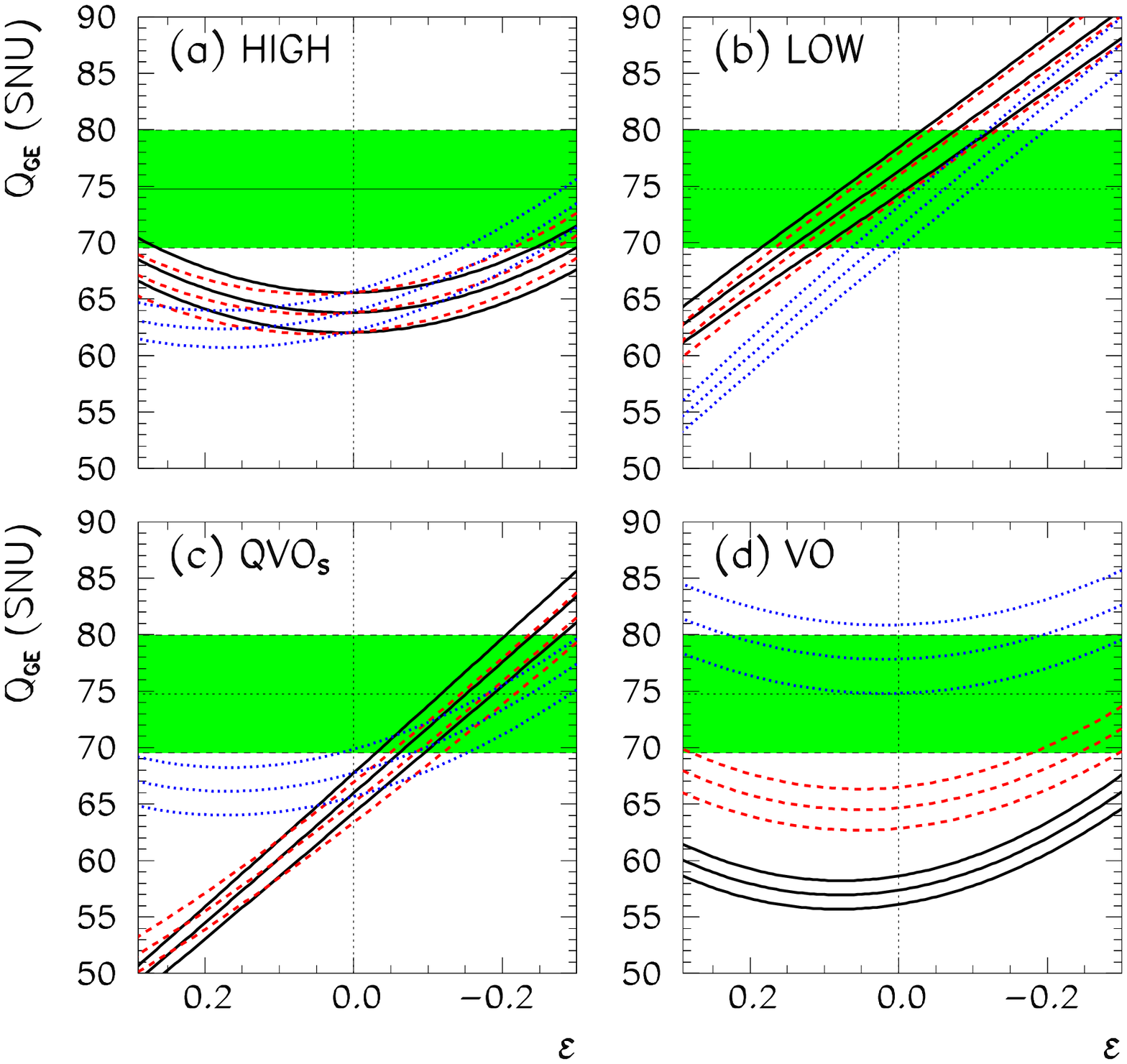,width=5.0in,angle=0}}
\tightenlines  
\caption[]{\small
The dependence of the Ge--production rate on $\epsilon$  for various 
values of $\Delta m^2$ within the (a) HIGH, 
(b) LOW, (c) QVO$_{{\rm S}}$ and (d) VO regions. 
Full, dashed and dotted lines correspond to: 
(a) $\Delta m^2=4,\ 0.8,$ and $0.2 \times 10^{-5}$ eV$^2$,    
(b) $\Delta m^2=1,\ 0.7,$ and $0.3 \times 10^{-7}$ eV$^2$,    
(c) $\Delta m^2=10,\ 5,$ and $1 \times 10^{-9}$ eV$^2$,    
(d) $\Delta m^2=5,.3,$ and $1 \times 10^{-10}$ eV$^2$, respectively.    
For each mass the three curves correspond to the predicted central 
value plus and minus 1$\sigma$ theoretical uncertainties.
The boron neutrino flux is normalized to the measured rate at 
Super--Kamiokande. The horizontal band corresponds to the averaged 
Gallium result and its 1$\sigma$ error. 
\label{q_ger}}
\end{figure}

Let us now study the implications of the results presented in 
Fig.~\ref{rates}(a) 
where we show the Ge--production rate as a function 
of $\Delta m^2$ for different values of the $\epsilon$ parameter
and in and Fig.~\ref{q_ger} 
where we plot $Q_{Ge}$ as a function of 
$\epsilon$ in the various $\Delta m^2$ regions.
While the results plotted in Fig.~\ref{rates}(a) are obtained within the
SSM, the predicted rates shown in  Fig.~\ref{q_ger} 
have been obtained after normalization of the 
boron neutrino flux to the Super--Kamiokande measured rate as described
previously for Fig.~\ref{q_ar}. In this case, unlike in the
case of Ar--production rate, the results are 
very slightly modified by the $^8B$ flux normalization 
since  the Ge--production rate is dominated by the
contribution from the $pp$ neutrino flux and the corresponding theoretical
uncertainties are smaller.   

In the QVO$_{{\rm L}}$ region, the rate is determined by averaged vacuum 
oscillations and therefore the expected rate is 
symmetric around maximal mixing.  
According to Eq.~(\ref{avprob2}), $P(\epsilon=\pm0.3)=0.545$ 
which corresponds to $Q_{Ge}=70.3\pm 3.5$ SNU (with $1 \sigma$
theoretical error). For exactly maximal mixing we have $P(\epsilon=0)=0.5$ 
and the rate is minimal, $Q_{Ge}=64.5\pm 3.5$ SNU. 
As seen in Fig.~\ref{rates}(a) the bands for the 
different $\epsilon$-values almost overlap and all the predictions are in  
agreement with the present experimental result. In other words, 
with present experimental error bars it is impossible to measure 
deviations from maximal mixing. 

In the part of the HIGH region with small $\Delta m^2$ the effects for  
$pp$ neutrinos occur in the transition between linear and quadratic
$\epsilon$ regimes. Consequently, the sensitivity of $Q_{Ge}$ to 
$\epsilon$ is still below the present experimental accuracy. We find 
from Fig.~\ref{rates}(a): 
$Q_{Ge}(\epsilon=0)\approx Q_{Ge}(\epsilon=+0.3) \approx 60-68$ SNU, 
while $Q_{Ge}(\epsilon=-0.3)\approx 70\pm 4$ SNU, approximately  
$1\sigma$ (experimental) higher. From Fig.~\ref{q_ger}(a)
we see that after the $^8B$ normalization the variation of 
the $Q_{Ge}$ in $\epsilon$ range $(-0.3,+0.3)$ is
$\Delta Q_{Ge}\sim(10 - 12)$ SNU, 
which is about $2\sigma$ (present experimental error).

In the LOW region the sensitivity to $\epsilon$ is maximal: 
the $pp$ neutrinos undergo  pure matter conversion and the rate 
depends on $\epsilon$ linearly. We get, for $\Delta m^2= 10^{-7}\ 
\mbox{\rm eV}^2$: $Q_{Ge}(\epsilon = - 0.3) = 93 \pm 4$ SNU,   
$Q_{Ge}(\epsilon = 0) = 77.5 \pm 3.5$ SNU and 
$Q_{Ge}(\epsilon = 0.3) = 61 \pm 3$ SNU. The difference between the 
predicted values of $Q_{Ge}(\epsilon=-0.3)$ and $Q_{Ge}(\epsilon=+0.3)$
is more than $2\sigma$ (experimental). From Fig.~\ref{q_ger}(b)
we see that after the $^8B$ normalization the variation of 
the $Q_{Ge}$ in $\epsilon$ range $(-0.3,+0.3)$ is
$\Delta Q_{Ge}\sim (33 - 35)$ SNU, 
which is more than $5\sigma$ (present experimental error). 
In this case one gets certain restrictions on $\epsilon$, although the
confidence level is low. For example, for $\Delta m^2= 10^{- 7}\ 
\mbox{\rm eV}^2$, 
the combined SAGE and GALLEX+GNO result gives the $1\sigma$ range 
$-0.12\leq\epsilon\leq+0.2$. 
Therefore, further decrease 
of the experimental error bars by a factor of two, from the present
5 SNU to 3 - 4 SNU, could have important implications for the mixing 
provided that $\Delta m^2$ will be fixed by some other independent 
measurement. 
Notice that in the LOW region one expects maximal regeneration effect for  
the $pp$-neutrinos which can be detected as, {\it e.g.}, seasonal variation 
of the  Ge--production rate~\cite{FLMP2}. 

The situation is similar in the QVO$_{{\rm S}}$ region down to 
$\Delta m^2 = 5 \times 10^{-9}\ \mbox{\rm eV}^2$: 
the Ge--production rate depends on $\epsilon$ 
linearly and $\Delta Q_{Ge} \sim 30$ SNU (see Fig.~\ref{q_ger}(c)). 

In the VO region, deviations from 
the maximal mixing result are determined by $\epsilon^2$  
and the variations (for a given $\Delta m^2$) are small as shown in
Fig.~\ref{q_ger}(d): 
$\Delta Q_{Ge} \sim 5 - 7$ SNU. Thus the sensitivity of present data is 
still low and practically the whole interval $\epsilon\in(- 0.3,+0.3)$ is
allowed at $(2 - 3) \sigma$ level.  

In consequence serious implications for maximal mixing require 
further decrease of the experimental error bars down to 3-4 SNU.

\subsection{The Day--Night asymmetry in electron scattering events}
\label{subsec:daynight}
In Fig.~\ref{ardn} we show contours of constant Day--Night asymmetry
of the $\nu-e$ scattering events in the $\Delta m^2 - \epsilon$ 
plane. The Super--Kamiokande 1117 days result
\begin{equation} 
A_{{\rm N-D}}=2 \frac{N-D}{N+D}\sim 0.034\pm 0.026\; 
\end{equation}
excludes at the $3\sigma$ level the region 
$A_{{\rm N-D}}\gsim 0.11$ which corresponds, at maximal mixing, to 
\begin{equation}
4 \times10^{-7}\lsim\Delta m^2\lsim 1.3\times 10^{-5}\ \mbox{\rm eV}^2.  
\label{dnexcl}
\end{equation}
The exclusion interval increases slightly with $\epsilon$. The preferable
regions of $\Delta m^2$ for $\epsilon = 0$ are 
\begin{equation}
\Delta m^2 = (2.5 - 10)\times 10^{-5}\  \mbox{\rm eV}^2, ~~~~ 
\Delta m^2 = (0.6 -2.2)\times10^{-7}\  \mbox{\rm eV}^2. 
\label{dnpref}
\end{equation}
We emphasize that these results are SSM independent and have no 
ambiguities related to the analysis of the data.

\subsection{Zenith angle distribution of electron scattering events}  
\label{subsec:zenith}

\begin{figure}[!t]
\centerline{\psfig{figure=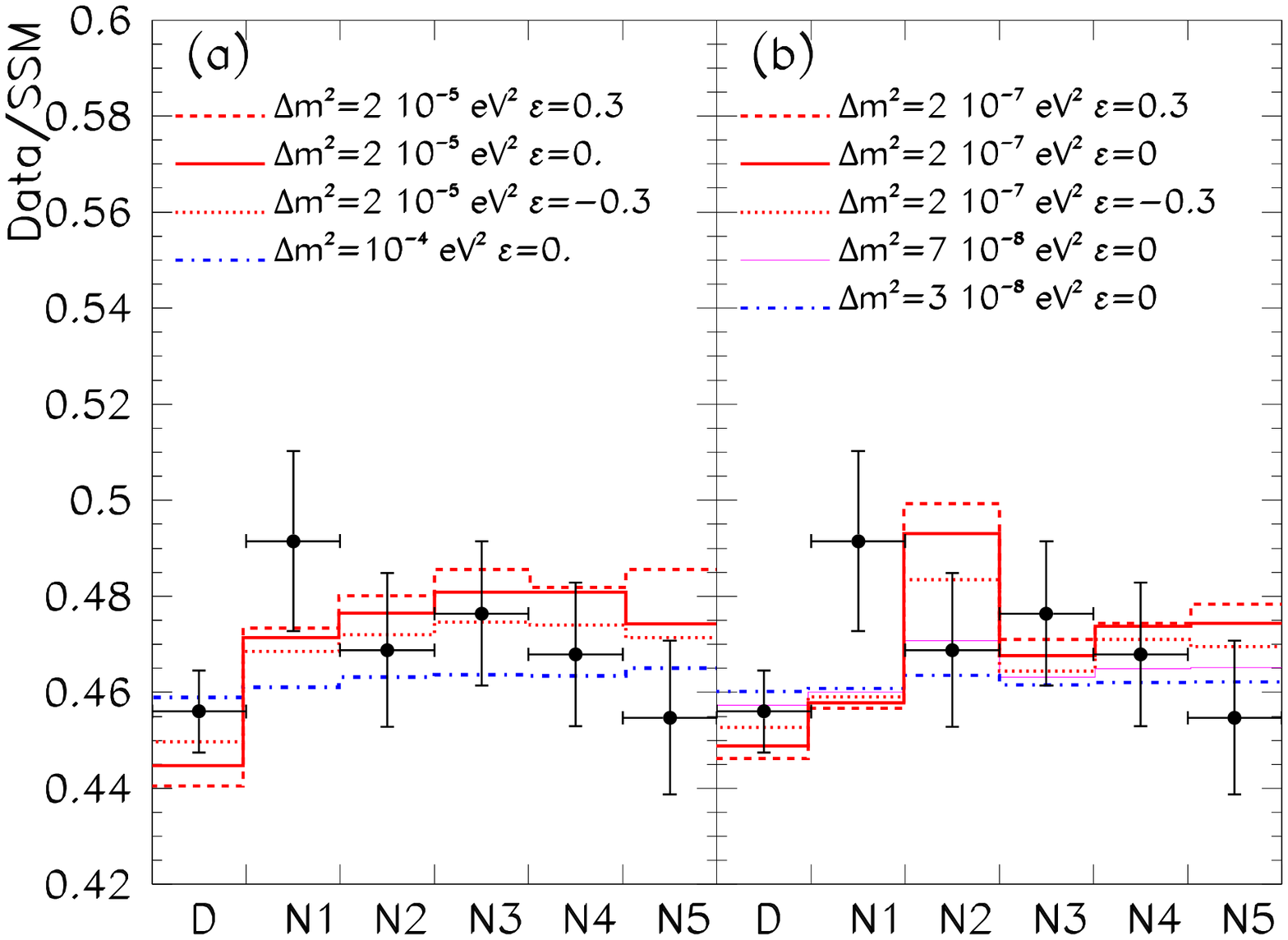,width=5.0in,angle=0}}
\tightenlines  
\caption[]{\small
The zenith angle distribution of events in Super--Kamiokande and the
predictions with (near--)maximal mixing for various 
values of $\Delta m^2$ within the (a) HIGH and (b) LOW  regions. 
\label{zenmax_new}}
\end{figure}

In Fig.~\ref{zenmax_new} we show the zenith angle distribution of 
events in Super--Kamiokande for maximal and near--maximal mixing
and for various values of $\Delta m^2$ from the allowed regions. 
Significant enhancement of the night rate is expected in the  
HIGH and LOW regions. 

In the HIGH region the distribution of events during the night is rather 
flat and the dependence on $\epsilon$ is weak, so it will be difficult to use 
the shape to measure $\epsilon$. The dependence on $\Delta m^2$ is also 
weak. Maximal signal is expected in the N3 or/and N5 (core) bins. 

In the LOW region the oscillations occur in the matter dominated regime 
(see Sec.~\ref{sec:physics}) when  the oscillation length practically 
coincides with the refraction length, $l_m \approx l_0$. 
For those trajectories crossing the mantle only  (N1-N4),  
the latter can be approximately determined by the average density along 
the trajectory. Maximal effect (which corresponds to the 
oscillation phase $=\pi$) is realized 
for the trajectories with $\cos \theta_Z = 0.3$,  {\it i.e.} in 
the N2 bin (see Fig.~\ref{zenmax_new}(b)). The phase $2\pi$ is collected at 
$\cos\theta_Z\sim 0.5$ which corresponds to a minimum of the signal.
Notice also that the zenith angle distribution depends on $\epsilon$ very 
weakly. For a given trajectory the amplitude of the oscillation is
determined by the mixing angle in the Earth matter 
\begin{equation} 
\sin^2 2\theta_E 
\sim \frac{1}{\eta_E^{-2} +1 - 2 \epsilon \eta_E^{-1}}   
\label{eta}  
\end{equation}
where $\eta_E$ the parameter $\eta$ in the Earth defined in Eq.~(\ref{etaE}). 
Thus $\sin^2 2\theta_E$ is an increasing function of $\epsilon$ 
(see also Eq.~(\ref{stseps})). As a consequence 
the number of events in maxima increases with $\epsilon$ as seen
in Fig.~\ref{zenmax_new}(b). Present data do not show any enhancement in
the N2 bin. 

We conclude that precise measurements of the zenith angle 
distribution would allow the determination of 
$\Delta m^2$ and probably resolve the HIGH/LOW ambiguity but are unlikely 
to play a significant role in the determination of $\epsilon$. 

\subsection{The recoil electron energy spectrum} 
\label{subsec:spectrum}

\begin{figure}[!t]
\centerline{\psfig{figure=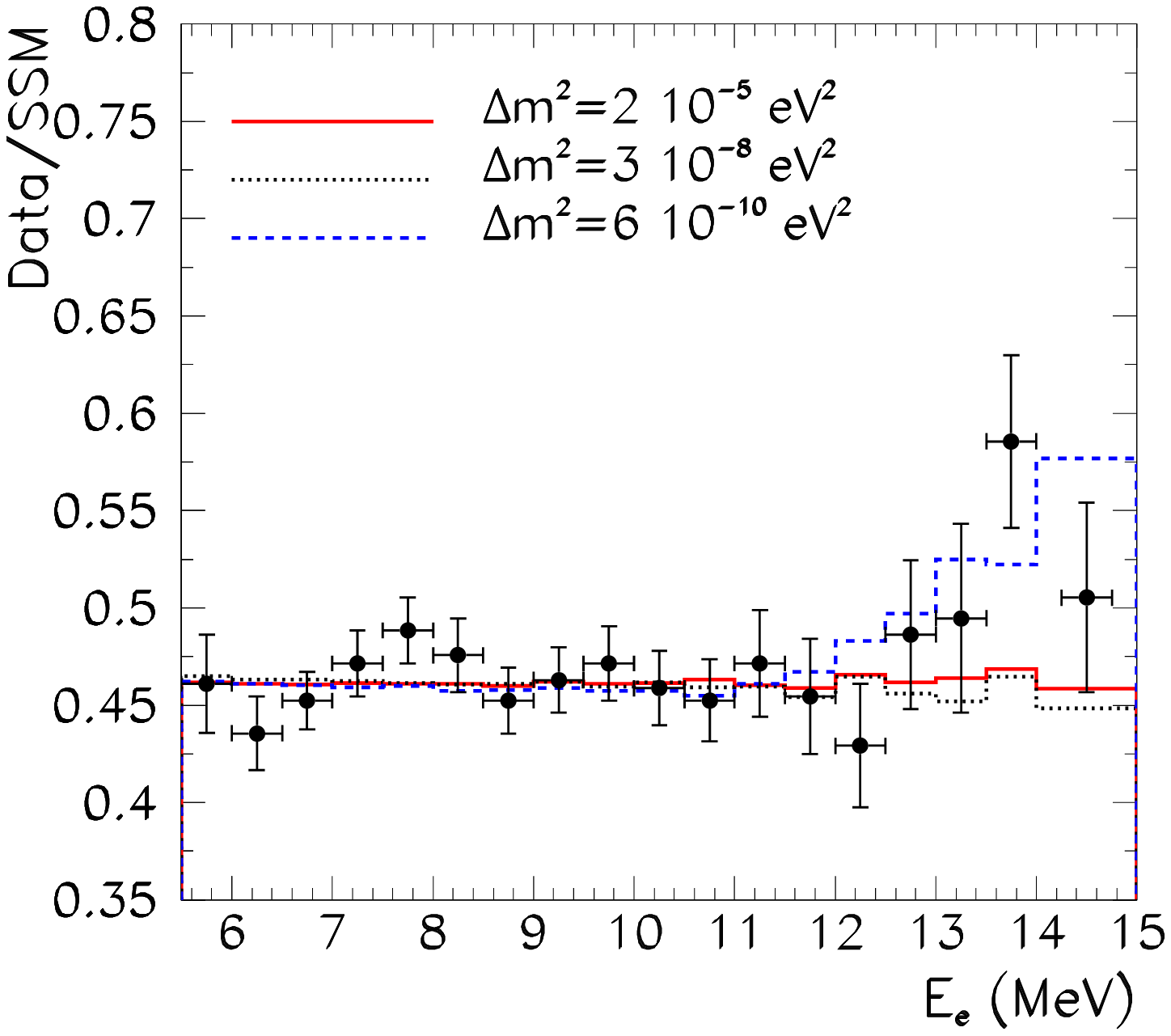,width=5.0in,angle=0}}
\tightenlines  
\caption[]{\small
The ``reduced'' spectrum for Super--Kamiokande for $\epsilon=0$ and various 
values of $\Delta m^2$. 
\label{specmax}}
\end{figure}

In Fig.~\ref{specmax} we show the expected recoil electron spectrum 
for maximal mixing with various values of $\Delta m^2$ in the
regions allowed by the global fit. In the HIGH and LOW regions the
``reduced" spectra,
$R(E)$, are flat. Strong distortion is expected in the VAC$_{\rm L}$
region. Thus further improvement on the measurement
of the recoil electron spectrum can discriminate between the MSW and 
VAC$_{\rm L}$ regions. 

\begin{figure}[!t]
\centerline{\psfig{figure=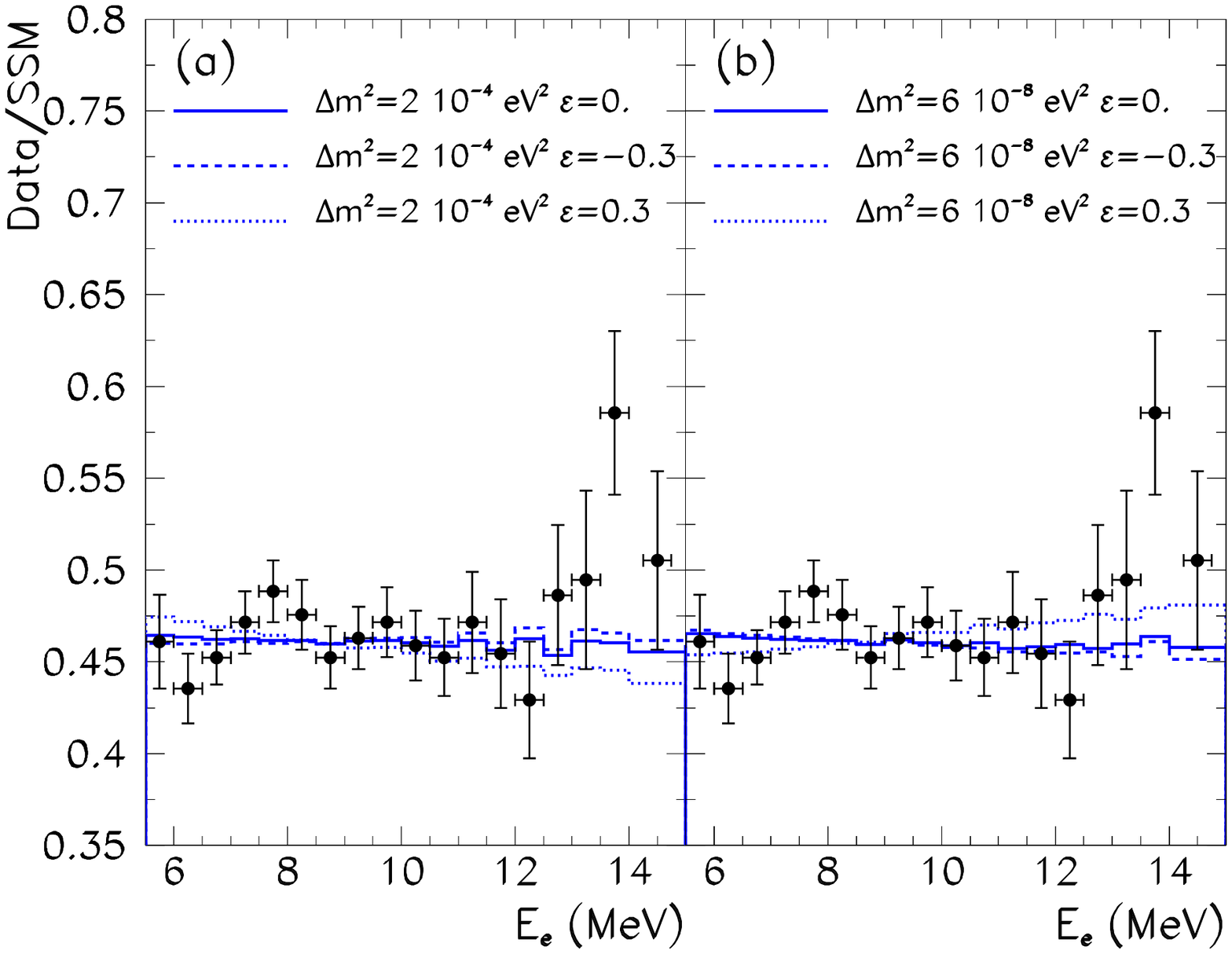,width=5.0in,angle=0}}
\tightenlines  
\caption[]{\small
The ``reduced'' spectrum for Super--Kamiokande for (a) 
$\Delta m^2=2\times 10^{-4}$ eV$^2$ and (b) $\Delta m^2=6\times 10^{-8}$
eV$^2$ with $\epsilon=-0.3,0,0.3$.
\label{specmax2}}
\end{figure}

For large enough $\Delta m^2$ and
$\epsilon \neq 0$ a distortion in the spectrum appears due to the effect of the
adiabatic edge (Fig.~\ref{specmax2}(a)). This feature can be
traced from the dependence of the survival probability on
$\Delta m^2/E$ (see Fig.~\ref{fig:sun}). For a positive
$\epsilon$ conversion inside the sun leads to negative slope
of the reduced spectrum: $R(E)$ is larger  at low energies.
For negative $\epsilon$,  $R(E)$ increases with $E$  and 
the slope is positive. 
In the small $\Delta m^2$  part of the LOW region
the distortion of the spectrum is induced by the effect of the
non-adiabatic edge (Fig.~\ref{fig:sun}, $\Delta m^2/4E = 
(0.3 - 3) \times 10^{-15}$ eV. Here the situation is opposite to
that for the HIGH region. The slope is positive for positive $\epsilon$
and negative for negative $\epsilon$ as illustrated in Fig.~\ref{specmax2}(b). 

As follows from Figs.~\ref{specmax} and~\ref{specmax2}, the measurement
of the electron energy spectrum provides information mainly on the 
value of  $\Delta m^2$ but 
it will be difficult to determine  $\epsilon$ by measuring
the slope (first moment of the spectrum) in the interval
$- 0.3 \leq \epsilon \leq  0.3$ at Super--Kamiokande. 
At present Super--Kamiokande has 
presented the measured value $\langle T\rangle=8.14 \pm 0.02$ MeV.
The precision of this measurement is to be compared with the maximum 
theoretically expected variation  
$\Delta \langle T\rangle=|\langle T\rangle(\epsilon=0.3)-
\langle T\rangle(\epsilon=-0.3)|<0.025$ MeV which occurs for the two 
values of $\Delta m^2$ shown in Fig.~\ref{specmax2}. Thus with the
existing experimental precision, in the MSW region, the full range of 
$\epsilon$ is allowed at $\sim$ 1$\sigma$.


\section{Tests of Maximal Mixing in Future Experiments}
\label{sec:future}

In this section we consider the prospects of testing (near--)maximal 
mixing  of the electron neutrinos in future experiments. We study various 
possibilities to measure the deviation $\epsilon$ and we estimate the
accuracy at which 
maximal mixing can be established or excluded. 
The extent to which the results of this Section hold in a three 
generation framework with $|U_{e3}|\neq0$ is discussed in 
Sec.~\ref{sec:threenus}.

\subsection{General requirements}
\label{subsec:requirements}
There are two requirements for a precise determination of the mixing: 

1. Uncertainties in the original neutrino fluxes should not play a role. 
For this purpose we will consider SSM independent observables, or at least 
observables which do not depend on the uncertainties in the boron neutrino
flux. 

2. At least two independent observables should be measured. 
As we have seen in Sec.~\ref{sec:status}, the survival probability
$P_{ee}$ and consequently all observables depend on both $\epsilon$ 
and $\Delta m^2$. Even in the case of maximal mixing when the solar 
survival probability is constant, $P_D=1/2$, a dependence of $P_{ee}$ 
on $\Delta m^2$ appears due to the Earth regeneration effect. 

Thus, to determine the mixing, one should find two independent observables 
which (i) are free of flux uncertainties, (ii) can be measured with 
high accuracy, and (iii) depend on different combinations of $\epsilon$ 
and $\Delta m^2$. In what follows we will identify such observables and 
study the accuracy at which mixing can be measured. 

\subsection{GNO and Super--Kamiokande}
\label{subsec:now}
The main objective of the GNO experiment \cite{gno} is to reach an accuracy 
$\sim 3$ SNU in the measurement of the Ge--production rate, $Q_{Ge}$.
Also seasonal variations of $Q_{Ge}$ will be studied. Super--Kamiokande 
will continue to collect data for at least 10 years. With an energy 
threshold as low as 5 MeV the accuracy in measuring the Day--Night 
asymmetry will improve to $\sim0.010-0.015$. 

Notice that, in the MSW region, $A_{\rm N-D}$ is mostly sensitive to 
$\Delta m^2$, whereas $Q_{Ge}$ has strong dependence on $\epsilon$. 
Therefore the pair of observables 
($A_{{\rm N-D}},Q_{Ge}$) is, in principle, 
very useful for measurements of the oscillation parameters in the matter 
conversion region. 

\begin{figure}[!t]
\centerline{\psfig{figure=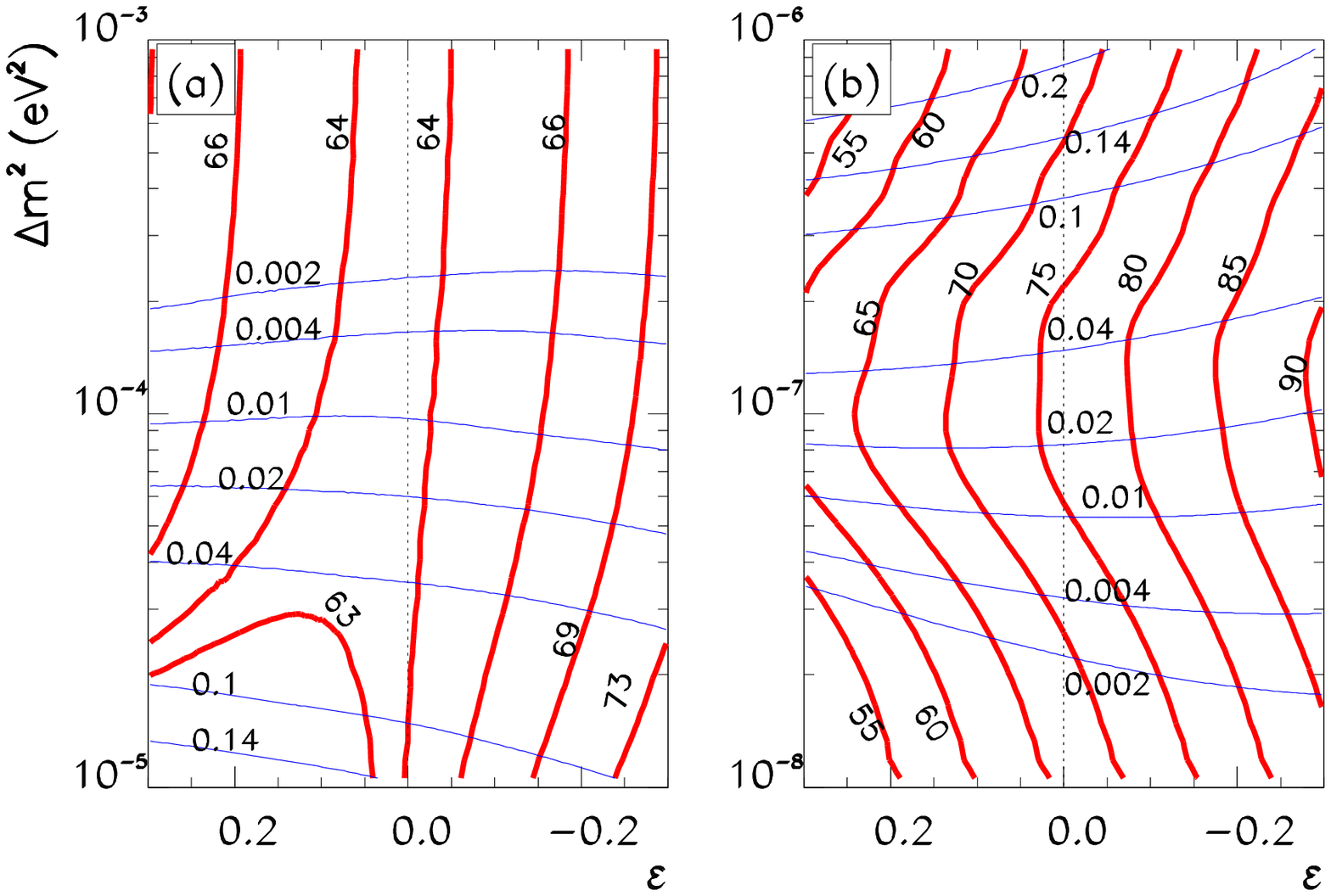,width=5.0in,angle=0}}
\tightenlines  
\caption[]{\small
Iso--contours of $A_{{\rm N-D}}$ at Super-Kamiokande (thin lines) and of 
$Q_{Ge}$ (thick curves with numbers in SNU) in the 
$\Delta m^2-\epsilon$ plane for the HIGH (a) and LOW (b) regions. 
\label{gerdn}}
\end{figure}

In Fig.~\ref{gerdn} we show simultaneously contours of constant 
$A_{\rm N-D}$ at Super--Kamiokande site and $Q_{Ge}$ in the 
$\Delta m^2-\epsilon$ plane. The
iso--contours for $Q_{Ge}$ have been obtained for central values of the 
solar fluxes according to BP98. The theoretical (1$\sigma$)
uncertainty is about $\pm 2$ SNU. 

The iso-contours of $Q_{Ge}$ and $A_{{\rm N-D}}$ are nearly perpendicular
to each other, which reflects the fact that these observables depend 
on different combinations of the oscillation parameters. However, 
the accuracy of the present experimental results  is not enough
to put statistically significant bounds on the mixing. The present
experimental $1 \sigma$ intervals are $Q_{Ge} \sim 69 - 80$ SNU 
and $A_{\rm N-D} \sim  0.01 - 0.06$. The resulting constraints
on the mixing parameters can be read from Fig.~\ref{gerdn}:
in the HIGH region, $-0.2\gsim\epsilon$ and 
$\Delta m^2\sim(1.5 - 8)\times 10^{- 5}\ \mbox{\rm eV}^2$, and in the LOW
region
$-0.10\lsim\epsilon\lsim+0.15$ and $\Delta m^2\sim(0.5-2)\times10^{- 7}\
\mbox{\rm eV}^2$. Inclusion of the theoretical uncertainties will 
expand these regions, especially in the HIGH case where the 
sensitivity of $Q_{Ge}$ to $\epsilon$ is rather low. 

Notice that the present Gallium data somewhat disfavor maximal mixing 
in the HIGH region. We estimate that the  
$1\sigma$ accuracy of the determination of $\epsilon$ is 
\begin{equation}
\Delta \epsilon \approx 0.15 - 0.20,  ~~~~ (1 \sigma). 
\label{accur1}
\end{equation}
Within $2 \sigma$ experimental errors the allowed regions cover 
most of HIGH parameter space of Fig.~\ref{gerdn}(a), and  
practically the entire LOW parameter space of Fig.~\ref{gerdn}(b). 
All values of
$\epsilon\in(- 0.3,+0.3)$ become allowed at the $2 \sigma$ level. 

With more data from GNO and higher statistics Super--Kamiokande measurements 
of the asymmetry one can reach better sensitivity. 

\subsection{SNO}
\label{subsec:sno}

The SNO experiment \cite{sno} will study various observables in 
three types of processes: 
\begin{enumerate}
\item Charged current neutrino-deuteron scattering:
the total rate above a certain threshold (we denote the reduced 
rate of events  by [CC]), the energy distribution of events (electron
energy spectrum), 
and the time variation of events which can be characterized by the Day-Night 
asymmetry $A_{\rm N-D}^{\rm CC}$, the zenith angle distribution, 
and the seasonal asymmetry.  
\item Neutrino-electron scattering: the total rate [ES], 
electron energy spectrum, and time variations. 
\item Neutral current neutrino-deuteron scattering: the total rate 
[NC],  and time variations. 
\end{enumerate}
Since the SNO observables depend on the boron neutrino flux only 
(we neglect the effect of the hep-neutrino flux), 
the ratios of rates are flux independent. Also  relative time
variations and energy spectrum distortion are flux independent.  
In what follows all the results will be presented for an energy
threshold of 5 MeV.
\begin{figure}[!t]
\centerline{\psfig{figure=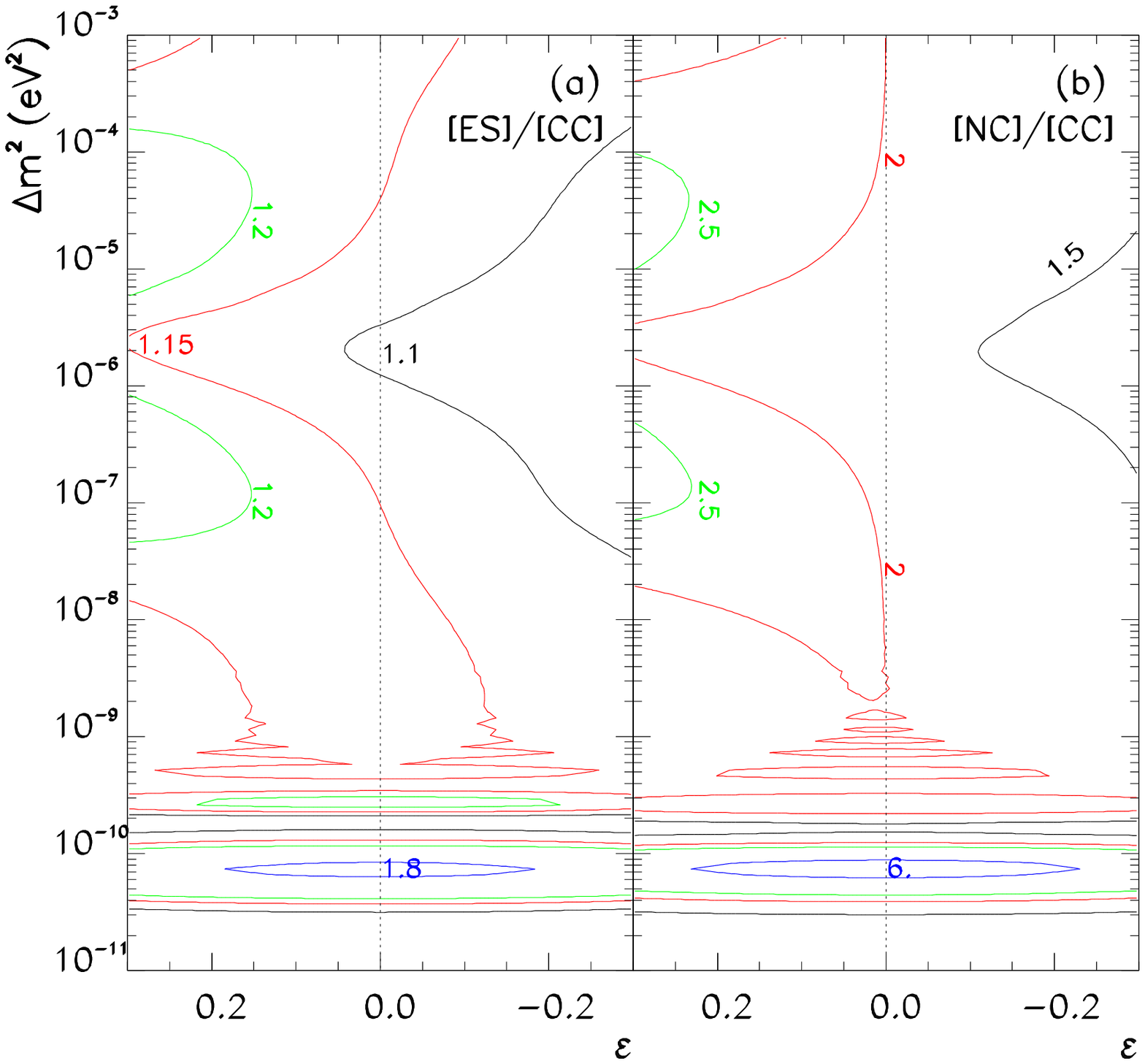,width=5.0in,angle=0}}
\tightenlines  
\caption[]{\small
Iso-contours of the double ratios of rates in the $\Delta m^2-\epsilon$ 
plane: (a) [ES]/[CC]; (b) [NC]/[CC] for SNO. 
\label{sno_rates}}
\end{figure}  

In Fig.~\ref{sno_rates}(a) we show the iso-contours of the double ratio 
[ES]/[CC] in the $\Delta m^2-\epsilon$ plane. In terms of averaged survival
probability,  $P(\Delta m^2,\epsilon)$, the ratio can be written  as  
\begin{equation}
\frac{\mbox{\rm [ES]}}
{\mbox{\rm [CC]}}\sim 1-r+\frac{r}{P}=1-r+\frac{2r}{f(1-\epsilon)}. 
\label{ESCC}  
\end{equation}
The last equality applies in the MSW region. The factor $f={\cal O}(1)$ 
and deviates from unity due to Earth matter effects which are small 
(see Eq.~(\ref{freglim})) and $r$ is the ratio between the $\nu_x-e$ and
$\nu_e-e$ (ES) cross sections. The ratio [ES]/[CC] is larger than 1 and 
depends rather weakly on the oscillation parameters. 
From Eq.~(\ref{ESCC}) we find the relation between the accuracy of 
measurements of [ES]/[CC], $\Delta (\mbox{\rm [ES]/[CC]})$ and the
corresponding accuracy of determination of $\epsilon$: 
\begin{equation}
\Delta \epsilon \sim \frac{f}{2r} 
\Delta  \left(\frac {\mbox{\rm [ES]}}{\mbox{\rm [CC]}} \right). 
\label{accurESCC}
\end{equation}
That is, the accuracy  is lowered by factor $1/2r \approx 3$.
As follows from Fig.~\ref{sno_rates}(b), in the MSW region the ratio 
[ES]/[CC] increases with $\epsilon$ for fixed $\Delta m^2$. 
It varies within the limits [ES]/[CC]$\approx1.15\pm0.10$ for 
$\epsilon=0.0\pm0.3$. This variation is comparable with the 
expected $1\sigma$ error which is dominated by the 
uncertainty in the neutrino-deuteron  cross-section 
($\sim 6\%$)~ \cite{crossdeu,bkssno} (statistical error  has been calculated 
assuming  5000 CC and 500 ES events). 
Consequently  no significant constraints on the oscillation parameters 
can be obtained, unless the uncertainty in the cross-section is reduced. 
According to Eq.~(\ref{accurESCC}), 10\% precision in [ES]/[CC] 
corresponds to $\Delta \epsilon \approx 0.3$.   

In Fig.~\ref{sno_rates}(b) we show the iso-contours of the double ratio 
[NC]/[CC] in the $\Delta m^2-\epsilon$ plane. In terms of 
$P(\Delta m^2,\epsilon)$, the ratio can be written  as 
\begin{equation} 
\frac{\mbox {\rm [NC]}}{\mbox{\rm [CC]}} 
\sim \frac{1}{P}  = \frac{2}{f(1 - \epsilon)}~, 
\label{NCCC}  
\end{equation} 
where, again, the last equality is valid in the MSW region. 
From Eq.~(\ref{NCCC}) 
we get for the accuracy in the determination of $\epsilon$ in the MSW region: 
\begin{equation}
\Delta \epsilon \sim \frac{f}{2}
\Delta  \left(\frac {\mbox{\rm [NC]}}{\mbox{\rm [CC]}} \right).
\label{accurNCCC}
\end{equation}
Here the prefactor is smaller than 1. Moreover, the double ratio [NC]/[CC] 
will be determined with much better accuracy than [ES]/[CC]. The
theoretical uncertainties related to the neutrino-deuteron cross-sections  
essentially cancel out. The total $1\sigma$ error, which includes a 
statistical error for 5000 CC and 2000 NC events,  is about $3.6 \%$. 
According to Eq.~(\ref{accurNCCC}), this corresponds to 
$\Delta\epsilon\sim0.04$ for fixed $\Delta m^2$. 
As follows from Fig.~\ref{sno_rates}(a), in the MSW region the ratio 
[NC]/[CC] increases with $\epsilon$ for fixed $\Delta m^2$. 
It varies within the limits $1.2-2.7$ for $\epsilon$ in the range 
$(-0.3,+0.3)$. 
This variation is much larger than the expected $1 \sigma$ error, 
$\Delta(\mbox{\rm [NC]/[CC]})\sim 0.08$  (for [NC]/[CC] = 2). 

\begin{figure}[!t]
\centerline{\psfig{figure=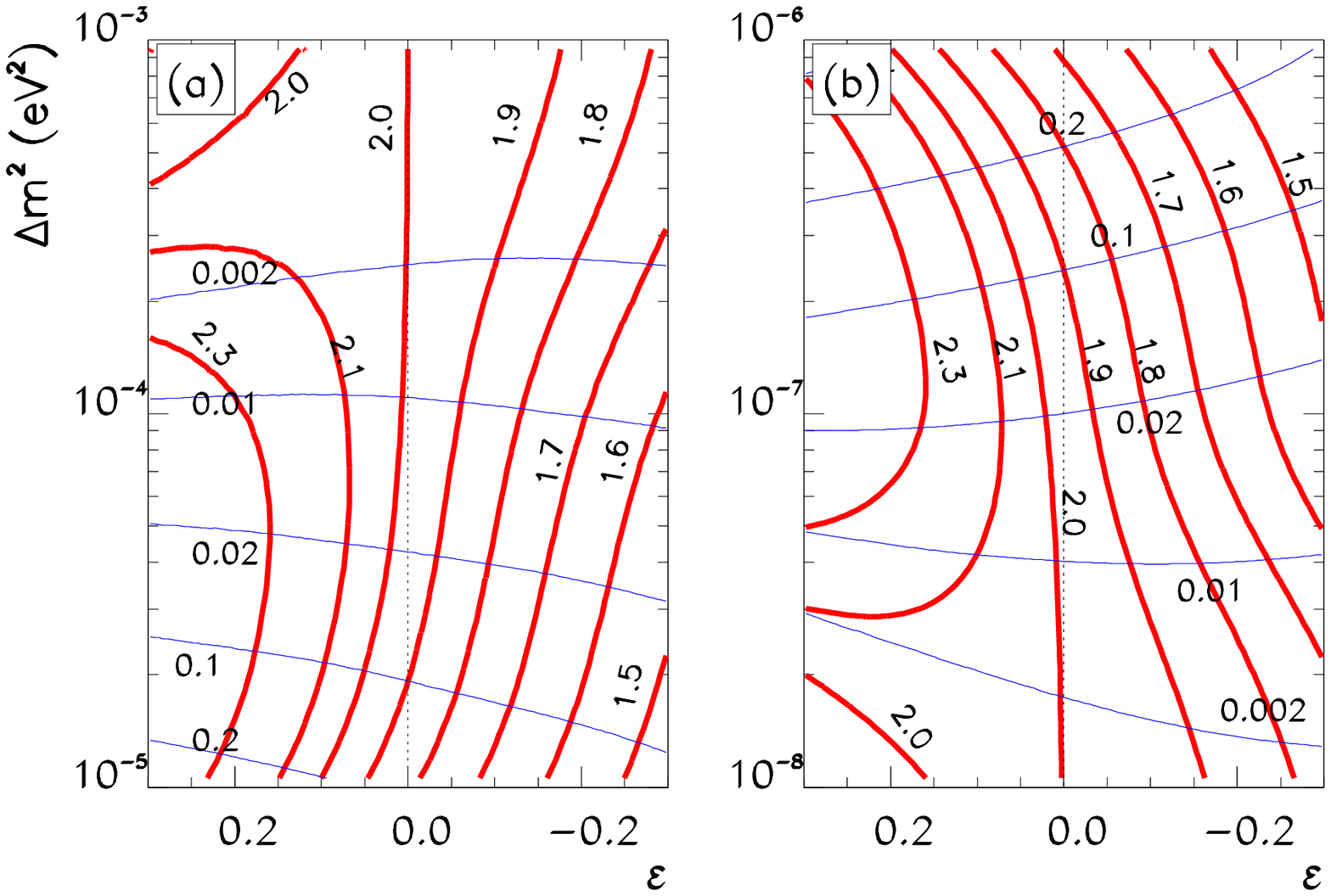,width=5.0in,angle=0}}
\tightenlines  
\caption[]{\small
Iso--contours of [NC]/[CC] (thick lines) and of $A_{{\rm N-D}}^{{\rm CC}}$ 
for SNO (thin lines) in the 
$\Delta m^2 - \epsilon$ plane in the (a) HIGH and (b) LOW regions.
\label{sno1}}
\end{figure}

In the allowed regions of the oscillation parameters, the ratio [NC]/[CC]  
depends strongly on $\epsilon$. A precise determination of $\Delta m^2$  
in these regions can be achieved from measurements of time variations, 
in particular, the Day- Night asymmetry 
$A_{\rm N-D}^{\rm CC}$. In
Fig.~\ref{sno1} we show iso--contours of 
$A_{\rm N-D}^{\rm CC}$ and [NC]/[CC] in the 
$\Delta m^2-\epsilon$ plane. Notice that the asymmetry of the CC-events 
is larger that the asymmetry of the ES-events for the same values of the
oscillation parameters. 
The contours have weak dependence on $\epsilon$. The combined analysis of 
[NC]/[CC] (sensitive mostly to $\epsilon$) and 
$A_{{\rm N-D}}^{{\rm CC}}$ 
(sensitive to $\Delta m^2$) can give a precise determination of the 
oscillation parameters. 
According to this figure, measurements yielding 
[NC]/[CC]$\sim2\times(1\pm0.04)$ and $A_{{\rm N-D}}^{{\rm CC}}\sim
0.1\times(1\pm0.3)$  will determine $\epsilon$ to an accuracy of order 
\begin{equation} 
\Delta \epsilon \approx 0.05 - 0.07~~~~~(1 \sigma)~.    
\label{accurNCdn}  
\end{equation}
Notice that the same pairs of values 
($A_{{\rm N-D}}^{{\rm CC}}$, [NC]/[CC]) 
appear in the HIGH and LOW regions. The LOW/HIGH ambiguity can be resolved
by the KamLAND \cite{kamland} reactor experiment which will give a 
positive oscillation 
signal in the case of the HIGH solution. It can also be resolved by the 
Borexino experiment which will show  strong  earth regeneration effect 
in the LOW region as discussed next. 

\subsection{Borexino}
\label{subsec:borex}
The Borexino collaboration \cite{borex} will measure the total rate of  
ES events and search for time variations of the signal.  

\begin{figure}[!t]
\centerline{\psfig{figure=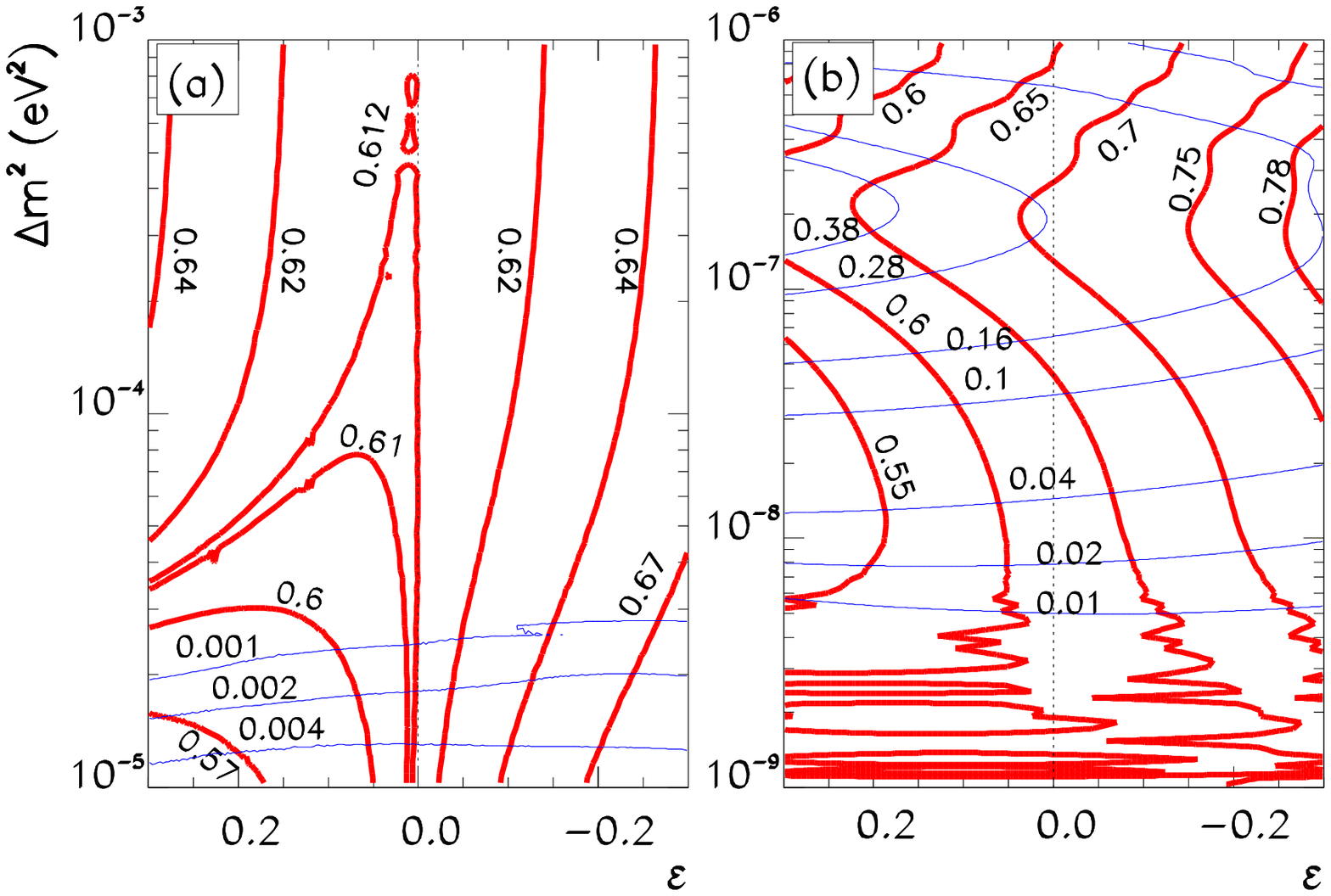,width=5.0in,angle=0}}
\tightenlines  
\caption[]{\small
Iso--contours of the reduced rate (thick lines) and the Day--Night asymmetry 
(thin lines) predicted for the Borexino experiment in the 
$\Delta m^2-\epsilon$ plane
for the QVO$_{\rm L}$ and HIGH (a), and LOW and QVO$_{{\rm S}}$ (b) regions. 
\label{borex}}
\end{figure}  

In Fig.~\ref{borex} we show iso-contours of the reduced rate $R_{Be}$ 
(suppression factor) and iso-contours of the Day-Night asymmetry  
in the $\Delta m^2-\epsilon$ plane. As discussed before for Super-Kamiokande
and SNO, the Day-Night asymmetry is sensitive mainly to $\Delta m^2$
while the deviation from maximal mixing can be restricted by the rate for 
which we can write
\begin{equation}
R_{Be} = r_{Be} + (1 - r_{Be}) P~, 
\label{ratebe}  
\end{equation}
where $r_{Be} \approx 0.24$ is the ratio of the $\nu_{\mu}-e$ to $\nu_{e}-e$ 
cross-sections for the beryllium neutrinos. 

From Eq.~(\ref{ratebe}) and Fig.~\ref{borex} we observe the following
behaviours:

(i) In the QVO$_{{\rm L}}$ region, the survival probability $P$ depends 
quadratically on $\epsilon$. We find $R_{Be}\sim0.62$ with very
weak $\epsilon$ dependence. 

(ii) In the HIGH region, the transition between the quadratic and linear 
$\epsilon$ dependence occurs. For $\Delta m^2 \sim 10^{-5}$ eV$^2$ 
the rate increases from 0.54 to  0.71 when $\epsilon$ decreases from 
$+0.3$ to $-0.3$. The Day--Night asymmetry is very small here. 

(iii) In contrast, in the LOW region Borexino has higher sensitivity to  
the oscillation parameters. For maximal mixing, $R_{Be}$ is in the interval 
0.65 - 0.70, and the asymmetry can be as large  as 30\%.  
The probability depends linearly on $\epsilon$, so that we get from 
Eq.~(\ref{ratebe}) 
\begin{equation}
\Delta \epsilon \sim \frac{2}{1 - r_{Be}} \Delta R_{Be}.  
\label{deltabe}
\end{equation}
Still, the sensitivity to $\epsilon$ is low but Borexino will play 
a crucial role in fixing $\Delta m^2$ and in particular by the
measurement of the Day--Night asymmetry  will be able to resolve
the HIGH/LOW ambiguity which may remain after the measurements at
Super--Kamiokande and SNO.

\subsection{LowNu Experiments}
\label{subsec:lownu}
A new generation of experiments aiming at a high precision real time 
measurement of the low energy solar neutrino spectrum is now under study
\cite{lowenergy}. Some of them such us HELLAZ, HERON and SUPER--MuNu 
\cite{le:nuscat} intend to detect the elastic scattering of the 
electron neutrinos with the electrons of a gas and  measure the recoil 
electron energy and its direction of emission. The proposed experiment 
LENS plans to detect the electron neutrino via its absorption in a 
heavy nuclear target with the subsequent emission of an electron and a 
delayed gamma emission \cite{le:abs}. 
The expected rates at those experiments for the proposed detector sizes 
are of the order of $\sim 1$--$10$ $pp$ neutrinos 
a day. Consequently, with a running time of two years  they can reach a 
sensitivity of a few percent in the total neutrino rate at low energy, 
provided that they can achieve sufficient background rejection. 
This would allow the determination of $\epsilon$ with a similar 
precision of a few percent,  in particular in the
QVO$_{\rm S}$, where SNO and Borexino cannot give much information due 
to their higher energy threshold.

Sensitivity to the oscillation parameters in the LMA region can also
be achieved in experiments detecting low energy $\bar\nu_e$ 
fluxes from nuclear reactors (detection threshold is about $E_{th}=1.8$ MeV). 
The Borexino experiment, in addition to detecting
solar neutrinos, aims to detect the diffuse fluxes from nuclear 
reactors in Europe, 
mainly in France and Germany, at an average distance $\sim 800$ km. 
The KamLAND experiment \cite{kamland} aims at detecting the low 
energy diffuse $\bar\nu_e$ fluxes from reactors in Japan from an average
distance of $\sim 200$ km. Both experiments 
can provide important information in discriminating between the HIGH and 
LOW solutions. However, they are expected to be very weakly sensitive
to $\epsilon$.  Given the short distances traveled by the neutrinos, 
matter effects are negligible for both experiments. Consequently, the  
survival probability for these experiments takes the vacuum form
of Eqs.~(\ref{pdepsL}) and ~(\ref{avprob2}) 
which depend only quadratically on $\epsilon$.  
With the expected achievable limits on the survival probability of
about $\sim 10$--20\%, a measurement of $|\epsilon|<0.3$ seems 
unfeasible.

\section{Effects of the third neutrino} 
\label{sec:threenus}

In the general case of three neutrino mixing when $|U_{e3}| \neq 0$ the 
definition of deviation from maximal mixing becomes ambiguous. 
Formally, maximal mixing of the $\nu_e$ in the three neutrino context 
can be defined as the equality $\theta_{12} = \pi/4$ where $\theta_{12}$
is the rotation angle in the plane of the first and second mass 
eigenstates. Phenomenologically, deviations from maximal mixing can be 
defined as a deviation of the averaged probability from 1/2 or 
a deviation of the depth of oscillations from 1.

Let us  consider neutrino mass spectrum which explains also the 
atmospheric neutrino problem via (mainly) $\nu_\mu-\nu_\tau$ 
oscillations. This implies
\begin{equation}
|\Delta m^2_{\odot}| \equiv |\Delta m^2_{21}| \ll
|\Delta m^2_{31}| \simeq |\Delta m^2_{32}| = 
|\Delta m^2_{{\rm atm}}|.
\label{hiedms1}
\end{equation}
In this case (as far as solar neutrinos are concerned) 
the oscillations driven by $\Delta m^2_{\rm atm}$ can be averaged.
This situation justifies our use of $\eta_S$ as defined in Eq. (\ref{cos}) 
and $\eta_E$ as defined in Eq. (\ref{etaE}) 
in the calculation of the survival probabilities.
We get the following expression for the survival probability \cite{cslim}: 
\begin{equation}
P_{e e} = (1 - |U_{e3}|^2)^2 P_2 + |U_{e3}|^4~, 
\label{Penote1}
\end{equation}
where $P_{2} = P(\Delta m^2, \theta, V) $ is the two neutrino mixing
survival probability 
determined by 
\begin{equation}
\Delta m^2 = \Delta m^2_{\odot}, ~~~~ \tan \theta = |U_{e2}/U_{e1}|, 
~~~~~V = V_{0} (1-|U_{e3}|^2).
\label{param}
\end{equation} 
where $V_0$ is the matter potential $V_0=\sqrt2 G_F\rho Y_e/m_N$.
In principle we can define a parameter $\epsilon_2$ which describes
the deviation from maximal mixing of $\theta$ defined in 
Eq.~(\ref{param}) similarly to the two neutrino parameters
$\epsilon$ and $\theta_{ex}$ of Eq.~(\ref{defeps}). Then 
the deviation from maximal mixing in the three neutrino context 
will depend on the specific physical situation. If the mass splitting 
$\Delta m^2_{\odot}$ and the mixing angle $\theta$ induce vacuum 
oscillations, then $P_2\approx(1 + \epsilon_2^2)/2$ and 
Eq.~(\ref{Penote1}) gives
\begin{equation}
P_{ee} \approx \frac{1}{2} + \frac{\epsilon_2^2}{2} - |U_{e3}|^2.  
\label{Penote2}
\end{equation}
Thus one can define the deviation from maximal mixing in the three 
neutrino context in this case as $\epsilon_3^v=\epsilon_2^2-2|U_{e3}|^2$. 
Clearly, the three neutrino case reduces to the two neutrino case if 
\begin{equation}
|U_{e3}|^2 \ll \frac{1}{2} \epsilon_2^2 .  
\label{ineq}
\end{equation} 
Taking $|U_{e3}|^2$ at the level of the CHOOZ bound \cite{CHOOZ},  
$|U_{e3}|^2 \sim 0.05$, Eq.~(\ref{ineq}) gives $|\epsilon_2| \gg 0.3$. 

If $\Delta m^2_{\odot}$ and $\theta$ lead to the adiabatic conversion 
in matter, then $P_2\approx(1-\epsilon_2)/2$ and Eq.~(\ref{Penote1}) 
gives 
\begin{equation}
P_{ee} \approx \frac{1}{2} - \frac{\epsilon_2}{2} - |U_{e3}|^2.
\label{Penote3}
\end{equation}
Here it is useful to characterize the deviation from maximal mixing
through $\epsilon_3^m=\epsilon_2+2|U_{e3}|^2$. Corrections due to the 
third generation can be neglected provided that 
\begin{equation}
|U_{e3}|^2 \ll \frac{1}{2} \epsilon_2.
\label{ineq1}
\end{equation}
For $|U_{e3}|^2 \sim 0.05$, Eq.~(\ref{ineq1}) gives $\epsilon_2\gg0.1$. 

We have verified that, in the case of adiabatic transitions, the results
of our calculations of the expected rates in the two-neutrino mixing
scenario can be translated to a good approximation to the case of 
three-neutrino mixing with the simple replacement of $\epsilon_2$ with
$\epsilon_3^m$. 
This applies, for instance, to the contours for the Ar--production rate 
in Fig.~\ref{ardn} and the predictions in Fig.~\ref{rates} and  
Fig.~\ref{q_ar} in the range $10^{-5}\gtrsim \Delta m^2$/eV$^2
\gtrsim 10^{-8}$. It also applies  
to the Ge--production rate in Fig.~\ref{gerdn}(b) and the corresponding
predictions in Fig.~\ref{rates} and Fig.~\ref{q_ger}(b) and~\ref{q_ger}(c). 
For the predictions of  
the SNO rates it can be used for Figs.~\ref{sno1} and~\ref{sno_rates} in the 
range  $10^{-4}\gtrsim \Delta m^2$/eV$^2\gtrsim 10^{-7}$, and
for the predicted rates at Borexino in Fig.~\ref{borex}(b) for 
$\Delta m^2$/eV$^2\gtrsim 5\times 10^{-9}$.

In the case of averaged vacuum oscillations (QVO$_{{\rm L}}$)
the results for the three-neutrino scenario  can be read from the results 
presented here with the replacement of $\epsilon_2^2$ with 
$\epsilon_3^v$. For long wavelength oscillations (QVO$_{{\rm S}}$ and VO), 
the results for the three-neutrino scenario cannot be 
directly derived from our results. Thus our predictions
in these regions only hold for very small values of the mixing angle 
$|U_{e3}|$, well below the present CHOOZ bound. 

Concerning the value of $U_{e3}$, although certain improvement 
on the present CHOOZ bound may be expected from long baseline
experiments, such as K2K and MINOS \cite{nufact1}, their final sensitivity
is still unclear as it depends on their capability of discriminating
against the $\nu_e$ beam contamination. Ultimate sensitivity can be
achieved at experiments performed with neutrino beams from muon-storage
rings at the so-called neutrino factories \cite{nufact2}.

\section{Maximal mixing and other experiments}
\label{sec:other}

In the previous sections we have concentrated on
effects in the solar neutrinos. Mixing of the electron neutrino can be
probed in a number of other experiments. 

\subsection{Atmospheric neutrinos}
\label{subsec:atmos}

Maximal and near--maximal mixing of the electron neutrinos can be
probed in the atmospheric neutrino studies. The oscillations in the 
Earth matter with parameters from the LMA or HIGH regions   
can give an observable effect in the $e$-like events.
The electron neutrino flux at the detector \cite{threeatm,yasuda,kim,PS}
can be  written as \cite{PS}
\begin{equation}
F_e = F_e^0\left[1 + P_{e \mu} (r \cos^2 \theta_{23} - 1)\right]~,
\label{excess}
\end{equation}
where $r \equiv F_{\mu}^0/F_e^0$ is the ratio of the original
electron and muon neutrino fluxes,
$P_{e \mu} = P(\Delta m_{\odot}^2, \epsilon_2)$ is the two neutrino   
transition probability, and $\theta_{23}$ is the $\nu_{\mu} - \nu_{\tau}$
mixing responsible for the dominant mode of the atmospheric neutrino
oscillations.
The transition probability can be of order one at
$\Delta m^2 > 3 \times 10^{-4}$ eV$^2$. It decreases fast with
$\Delta m^2/E$ due to matter suppression of the mixing.
Thus the biggest effect is expected in the low energy (sub-GeV)
events sample. Notice that the probability $P_{e \mu}$ enters in 
Eq.~(\ref{excess})
with a ``screening factor" $(r \cos^2 \theta_{23} - 1)$
which turns out to be small. Indeed, for the sub-GeV sample $r \approx 2$
and the screening factor is exactly zero for maximal mixing 
in the atmospheric neutrinos \cite{PS,ahlu}. The factor equals approximately
$\cos 2 \theta_{23}$,  so that for $\sin^2 2 \theta_{23} = 0.95$
we get about 0.22.

For $\theta_{23} < \pi/4$ the oscillations lead to the excess
of the $e$-like events. Indeed some excess is hinted by
the SK  data. The excess can be defined as $N_e/N_e^0 -1$, where  
$N_e$ and $N_e^0$ are the numbers of events with and 
without oscillations. This excess can be written in a matter dominant
regime of oscillations ($\eta_E \ll 1$) as
\begin{equation}
\frac{N_e}{N_e^0} -1 \approx \cos 2 \theta_{23} \eta_E^2
(1 - \epsilon_2\, \eta_E),
\label{excess1}
\end{equation}
where $\epsilon_2$ was defined below Eq.~(\ref{param}).
The excess depends on $\epsilon_2$ linearly and it increases with
$\eta_E$. However it is even more sensitive to deviations of 
$\theta_{23}$ from the maximal mixing value.
For $\Delta m^2_{21} = 10^{-4}$ eV$^2$
and $\sin^2 2 \theta_{23} = 0.95$ the excess reaches about 3\%
and the dependence on $\epsilon$ is weak.
For $\Delta m^2_{21} =2 \cdot 10^{-4}$ eV$^2$ and the same $\theta_{23}$,
we find the excess  about 4.5 \%.
Significant dependence on $\epsilon$ appears for
$\Delta m^2_{12} = 3 \times 10^{-4}$ eV$^2$. 
However, it is unlikely given 
the size of the effect that atmospheric neutrino data will give any 
significant information on the value of $\epsilon_2$.

\subsection{Supernova  neutrinos}
\label{subsec:supernova}

Maximal or near--maximal mixing of the electron neutrinos
will significantly modify properties of neutrino bursts from supernovae.
The effects depend crucially on features of the whole neutrino mass
spectrum and in particular on the value of $U_{e3}$ and whether the
mass hierarchy is normal ($\Delta m^2_{31} > 0$)
or inverted  ($\Delta m^2_{31} < 0$). Let us summarize here the main results
(for more details see \cite{DS} and references therein).

All oscillation and conversion  effects in supernova neutrinos are 
determined by the total survival probability of the electron neutrinos 
which in this subsection we will write as $p$, and total survival 
probability of the electron antineutrinos, $\bar{p}$. (This property is 
related to the fact that the original spectra of the muon and the tau 
neutrinos are identical and that the muon and tau neutrinos cannot be 
distinguished at the detection point). 

The probabilities should include the effects of propagation inside the 
star, on the way to the Earth and inside the Earth. Using the 
probabilities $p$ and $\bar{p}$, one can write the fluxes 
of the electron neutrinos, $F_e$, and electron antineutrinos, $F_{\bar e}$,
at the detector in terms of the original electron (anti)neutrino fluxes,
$F_e^0$ and  $F_{\bar e}^0$,  and the non-electron neutrino flux 
$F_x^0 \equiv F_{\mu}^0=F_{\tau}^0=F_{\bar{\mu}}^0=F_{\bar{\tau}}^0$:
\begin{equation}
F_e = p \cdot F_e^0 +  (1 - p) \cdot  F_x^0~, ~~~~
F_{\bar e} = \bar{p} \cdot F_{\bar e}^0 +  (1 -\bar{p}) \cdot  F_x^0~.
\label{flux}
\end{equation}
In general, $p$, and $\bar{p}$ depend on the neutrino energy.

Let us summarize the results for specific neutrino mass and flavor
spectra.

1) If $|U_{e3}|^2 > 3 \times 10^{-4}$ the conversion in the resonance
related to the largest (atmospheric)  splitting ($\Delta m^2_{31}$)
will be completely adiabatic and the final effect depends on the type of
mass hierarchy. In the case of normal hierarchy
($\nu_3$ is the heaviest state) the resonance conversion occurs
in the neutrino channel and for the survival probability we get
\begin{equation}
p \approx  |U_{e3}|^2 \ll 1.
\label{sprobab}
\end{equation}
This probability is practically independent of
the properties (mass, flavour) of the first and second mass
eigenstates. In particular, there is no sensitivity to
$\epsilon$ and no Earth matter effect is expected for neutrinos.

In contrast, the  antineutrino channels will not be affected by the high
resonance and $\bar{p}$ will be determined by physics of the two light
levels.

For parameters in the HIGH and LOW regions, the neutrino propagation in the
star is adiabatic, so that the survival probability in the star equals
\begin{equation}
\bar{p} \approx \cos^2 \theta_{12} = \frac{1}{2}(1 + \epsilon).
\label{sprobab1}
\end{equation}
This probability can be further modified due to
oscillations in the matter of the Earth.
Thus, we expect the following consequences:
(i) disappearance of the $\nu_e$ neutronization peak; 
(ii) hard $\nu_e$ spectrum (coinciding with the original
$\nu_{\mu}$) spectrum at the cooling stage:
\begin{equation}
F_e \approx  F_x^0~;
\label{flux1}
\end{equation} 
(iii) composite $\bar{\nu}_e$ spectrum:
\begin{equation}
F_{\bar e} = \frac{1}{2}(F_{\bar e}^0 +  F_x^0) -
\frac{\epsilon}{2}(F_x^0 -  F_{\bar e}^0) ;
\label{flux3}
\end{equation}
(iv) strong Earth matter effect
(which leads to different signals at various detectors). 
For the HIGH mass range, the Earth effect is maximal in
the high energy part of the spectrum, $E > 20$ MeV, whereas for the 
LOW solution the largest effect  is in the low energy part.

According to Eq.~(\ref{flux3}), the $\epsilon$-dependent term is
proportional to  the difference of the original fluxes. Thus
due to the uncertainties in the predicticted fluxes it will be difficult 
to measure $\epsilon$. In order to reduce the theoretical
uncertainty, one could in principle compare numbers  
of $\nu_e$ and $\bar{\nu}_e$ events at large $E$ which are determined by,
respectively, $F_x^0$ and $F_x^0 (1 - \epsilon)/2$ and are proportional 
to the same flux.

For parameters of the two light states in the VO region, the neutrino 
propagation in the star is non-adiabatic, so that the survival probability
can be writen as 
\begin{equation}   
\bar{p} \approx (1 - P_c) \cos^2 \theta_{12}  +P_c \sin^2 \theta_{12} = 
\frac{1}{2} + \epsilon \left(\frac{1}{2} - P_c\right),
\label{sprobabvac}
\end{equation}  
where the jump probability $P_c$ depends on the
details of the density profile in the outer regions of the star and 
cannot be reliably predicted.  Practically, the probability should 
lie between the adiabatic value (\ref{sprobab1}) and the pure vacuum 
oscillation expression $p =(1 + \epsilon^2)/2$.

In the case of inverted mass hierarchy the sensitivity to
$\epsilon$ appears in the neutrino channel and
neutrinos and antineutrinos interchange their roles.
Now the resonance is in the antineutrino channel so that
\begin{equation}
\bar{p} \approx  |U_{e3}|^2 \ll 1,
\label{sprobabnu}
\end{equation}
and the oscillations in the neutrino channels will be determined by physics 
of the two light levels. For parameters from the HIGH and LOW regions 
the propagation in the star is adiabatic and
\begin{equation}
p \approx \sin^2 \theta_{12} = \frac{1}{2}(1 + \epsilon).
\label{sprobab2}
\end{equation}
Moreover, Earth matter effects are expected for neutrinos.
For the VO region we find, similarly to Eq.~(\ref{sprobabvac}), 
\begin{equation}
p \approx \frac{1}{2} - \epsilon \left(\frac{1}{2} - P_c\right).
\label{sprobabvac2}
\end{equation}

In this inverted scheme we predict: (i) partial disappearance
of the $\nu_e$ neutronization peak; (ii) hard spectrum of the
electron antineutrinos:
\begin{equation}
F_{\bar{e}} \approx  F_x^0~;
\label{fluxanti}
\end{equation}  
(iii) composite $\nu_e$ spectrum:
\begin{equation}
F_{e} = \frac{1}{2}(F_{e}^0 +  F_x^0) +
\frac{\epsilon}{2}(F_x^0 -  F_{e}^0);
\label{flux4}
\end{equation}
(iv) Earth matter effects are expected in the neutrino channel only.

2) If $|U_{e3}|^2 < 3 \times 10^{-6}$, the effect of the third neutrino
can be neglected: in the resonance channel the transition driven by $U_{e3}$
is strongly non-adiabatic, and in the non-resonance channel the mixing
is always very small. In this case the problem is reduced to transitions
in two level system with parameters determined by $
\Delta m^2_{21}=\Delta m^2_{\odot}$ and $\epsilon$.
As a result both neutrino and antineutrino channels turn out to be
sensitive to $\epsilon$. The effects include those considered above both
for normal and inverted mass hierarchy.

For HIGH and LOW regions of parameters the propagation proceeds
adiabaticaly, and for the survival probabilities we get the expressions
given in (\ref{sprobab1}) and (\ref{sprobab2}). Correspondingly,
neutrino and antineutrino spectra will be given by Eq.~(\ref{flux3}) 
and (\ref{flux4}).
Thus we predict that both neutrino and antineutrino  spectra
will be composite, consisting of nearly equal admixture
of the soft and  hard components.
In the high energy part where effects of the soft components can be
neglected we get from Eqs. (\ref{flux3}) and (\ref{flux4}):
\begin{equation}
\frac{F_{e}}{F_{\bar{e}}} \sim 1 +  2\epsilon .
\label{flux5}
\end{equation}
That is, larger $\nu_e$ signal (as compared with $\bar{\nu_e}$)
is expected for $\epsilon > 0$ and smaller for $\epsilon <  0$.
Also Earth matter effect is expected in both neutrino and antineutrino
channels.

3) If $|U_{e3}|^2$ is in the intermediate region,
$3 \times 10^{-6} -  3 \times 10^{-4}$, the adiabaticity
in high mass resonance  is partially violated and we expect some
intermediate situation between those described in 1) and 2).
In particular, both $\nu_e$ and $\bar{\nu}_e$ spectra will be
composite, however admixtures of the soft and  hard
components will be unequal, {\it etc}..

To conclude, one expects strong influence of maximal and near--maximal
mixing on the properties of the neutrino bursts. However, the  
uncertainties in the predicted neutrino spectra will make it
difficult to obtain high sensitivity to $\epsilon$. 

Notice also that the analysis of the SN1987A data gives the 99\% CL bound on
$p > 0.65$. This would correspond to 
$\bar{p} > 0.3$ \cite{SSB}. Some recent calculations show that 
the difference between $\bar{\nu}_e$ and $\bar{\nu}_{\mu}$ original spectra 
can be rather small, which would somewhat relax the above bound. 

\subsection{Neutrinoless double beta decay}
\label{subsec:double}

The effective Majorana mass of the electron neutrino $m_{ee}$
relevant for the neutrinoless double beta decay is sensitive to the
distribution of the electron neutrino flavor in the 
mass eigenstates (see \cite{doubleteo} for recent discussion). 
The contribution to the effective mass $m_{ee}$ from the two
mass eigenstates responsible for the solar neutrino conversion can be
written in terms of the oscillation parameters as
\begin{equation}
m_{ee} = \frac{1}{2} \left| m_1 (1 + \epsilon)
+ \sqrt{m_1^2 + \Delta m_{\odot}^2} (1 - \epsilon) e^{i\phi_{12}}\right|,
\label{majmass}
\end{equation}
where $m_1$ is the mass of the first eigenstate and $\phi_{12}$
is the relative phase of the first and the second mass eigenvalues.
In the case of strong mass hierarchy, $m_1^2 \ll \Delta m_{\odot}^2$
we get from Eq.~(\ref{majmass})
\begin{equation}
m_{ee} \approx \frac{1}{2} \sqrt{\Delta m_{\odot}^2} (1 - \epsilon).
\label{majmassh}
\end{equation}
According to this equation in the HIGH region the effective mass can
be as big as  $(1 - 2)\times 10^{-2}$ eV which can be probed at the next
generation of the double beta decay experiments \cite{doubleexp}.
Notice that the contribution from the third mass eigenstate is strongly
restricted by present experimental bound $m_{ee} < 2 \times 10^{-3}$ eV.

Although the dependence of $m_{ee}$ on $\epsilon$ is rather strong,
it will be difficult to measure $\epsilon$ due to the uncertainties in the
nuclear matrix elements.  Eq.~(\ref{majmassh}) can be considered as
a test equation: if the measured values of $m_{ee}$, $\epsilon$ and
$\Delta m_{\odot}^2$ indeed satisfy this equation (within
experimental and theoretical  uncertainties) it will testify for
the validity of whole scheme.

In the case of strong mass degeneracy, $m_1^2 \gg \Delta m_{\odot}^2$
we get
\begin{equation}
m_{ee} \approx \frac{1}{2}m_1
\left| (1 + \epsilon) +  (1 - \epsilon) e^{i\phi_{12}} \right|,
\label{majmassd}
\end{equation}
where both non-oscillation parameters $m_1$ and $\phi_{12}$ are unknown.
For  $\phi_{12} = \pi$ and $(0)$ the mass equals
$m_{ee} = m_1 \epsilon$ $(m_1)$, so that $\epsilon$
determines the lower bound on $m_{ee}$. If $m_{ee}$ and $\epsilon$
are measured, the above equalities will determine the upper bound
and the lower bounds on the absolute scale of the neutrinos mass:   
$m_{ee} < m_1 < m_{ee}/\epsilon$.

In the case of inverted mass hierarchy the two states responsible for the
solar neutrino conversion are degenerate:
$m_1 \approx m_2 \approx \sqrt{\Delta m^2_{\rm atm}}$  and the
effective majorana mass can be written as
\begin{equation}
m_{ee} \approx \frac{1}{2} \sqrt{\Delta m^2_{\rm atm}}
\left| (1 + \epsilon) +  (1 - \epsilon) e^{i\phi_{12}} \right|.
\label{majmassinv}
\end{equation}
In this case the measurement of $\epsilon$ will allow  us to
determine the phase $\phi_{12}$.
According to Eq.~(\ref{majmassinv}), $\sqrt{\Delta m^2_{\rm atm}}\epsilon
<m_{ee}<\sqrt{\Delta m^2_{\rm atm}}$
which can be used as a test inequality for a given scheme.

Thus measurements of $\epsilon$ in the oscillation experiments will allow
to determine or restrict the effective mass $m_{ee}$ in the context
of certain schemes of neutrino masses and mixing.

\section{Conclusions}
\label{sec:conclusion}

In this work we have explored the phenomenological consequences
of (near--)maximal mixing of electron neutrinos with other standard
neutrinos. The possibility of such maximal or near--maximal lepton 
mixing constitutes an intriguing challenge for fundamental theories
of flavour. Our aim was twofold. First we have formulated  the present
status of maximal mixing of $\nu_e$ in the light of existing experimental 
data from solar neutrino experiments. Second we have explored the best 
ways to measure deviations from such maximal mixing at future 
experiments. 

We show in Sec.~\ref{sec:physics} that both probabilities and observables 
depend on $\epsilon$ quadratically in the regions of $\Delta m^2$ where 
the effects are due to vacuum oscillations, and they depend on 
$\epsilon$ linearly when matter effects dominate. Consequently, 
for $|\epsilon|\ll1$ the highest sensitivity to deviation from 
maximal mixing can be achieved in the $\Delta m^2$ ranges of the MSW 
effect. 

The results of a global fit to the existing solar neutrino data are
presented in Sec.~\ref{sec:status} and summarized in Figs.~\ref{glob}
$-$\ref{chi2_part}. From this analysis we find that values of the
mixing parameter $|\epsilon|\equiv|1-2\sin^2\theta|<0.3 $ 
are allowed at 
99\% or lower  CL for $\Delta m^2\gsim 1.5\times  10 ^{-5}$ eV$^2$ 
(which contains the  HIGH and QVO$_{\rm L}$ regions) and for
$4\times 10^{-10}$ eV$^2 \lsim\Delta m^2\lsim 2\times 10 ^{-7}$ eV$^2$ 
(which contains the defined LOW, QVO$_{{\rm S}}$ and upper 
VAC$_{{\rm L}}$ regions). 

The role of the individual existing experiments on the determination
of these regions is discussed in Sec.~\ref{sec:predic}. We conclude that
the present sensitivity to the mixing angle arises from the measurements
of total event rates. The present data from Homestake experiment 
in Ar--production rate gives the strongest constraint on maximal or 
near--maximal mixing as it favours a significant deviation from $\epsilon=0$.
This conclusion is independent of the existing theoretical uncertainty 
on the boron flux as discussed in Sec.~\ref{subsec:argon}. 
On the other hand the measurement of the zenith angle dependence and the 
recoil electron energy spectrum are important in the determination of 
the allowed mass ranges but are very weakly sensitive to deviation from 
maximal mixing. With the present existing sensitivity all values of
$|\epsilon|<0.3$ are allowed within 4$\sigma$.

In Sec.~\ref{sec:future} we have discussed the ways to improve our 
knowledge on deviations from maximal mixing at future experiments. 
We concentrate on observables which are SSM (or at least boron flux) 
independent. In order to determine both the mass and the mixing we study 
pairs of observables. 
First we have looked at the maximal sensitivity which may be achievable  
on the presently running experiments GNO and Super--Kamiokande. 
In principle the measurement of the Ge--production rate at GNO and 
the Day--Night asymmetry at Super--Kamiokande and SNO can give crossed
information
on the oscillation parameters in the matter conversion region. In practice,
however, the expected sensitivity is not enough to substantially improve 
the present knowledge on $\Delta m^2$ and $\epsilon$. The role of
SNO and Borexino experiments is discussed in Secs.~\ref{subsec:sno} and
\ref{subsec:borex}. We show that 
with the expected theoretical and
statistical uncertainty the most sensitive observable
to the mixing angle is the rate [NC]/[CC] measurable at SNO.  
For instance, a measurement yielding 
[NC]/[CC] $\sim2\times(1\pm0.04)$ and $A_{{\rm N-D}}^{{\rm CC}}\sim
0.1\times(1\pm0.3)$  will determine $\epsilon$ to an accuracy of order
$\Delta \epsilon  \sim 0.07$.  There exist however an ambiguity on the
allowed mass range
between HIGH and LOW regions. We show that the LOW/HIGH ambiguity can be 
resolved by the measurement of the Day--Night asymmetry at Borexino
experiment which is sensitive to strong Earth regeneration effect in 
the LOW region or by the detection of oscillations in long baseline
reactor experiments such as KamLand. However no substantial improvement
on the knowledge of $\epsilon$ is expected neither from Borexino nor
from the new generation of low energy experiments either with solar or 
reactor neutrinos.

\acknowledgments
We thank J. N. Bahcall, P. I. Krastev and E. Lisi 
for valuable discussions. We are particularly
indebted to 
Y. Suzuki for providing us with Super--Kamiokande data on the night
and day spectra. 
YN is partially supported by the Department of Energy 
under contract No.~DE--FG02--90ER40542, by the Ambrose Monell Foundation, 
by AMIAS (Association of Members of the Institute for Advanced Study),  
by the Israel Science Foundation founded by the Israel Academy of Sciences
and Humanities, and by the Minerva Foundation (Munich).  
AYS acknowledge partial support from NSF grant
No. PHY95-13835 to the Institute for Advanced Study.  
This work was also supported by the spanish DGICYT under grants PB98-0693 
and PB97-1261, by the Generalitat Valenciana under grant
GV99-3-1-01 and by the TMR network grant ERBFMRXCT960090 of the 
European Union.



\end{document}